\documentclass[preprint]{aastex}
\usepackage{graphicx}
\usepackage[section]{placeins}
 \usepackage{url}
\usepackage{amsmath}
\usepackage{dsfont}
\usepackage[toc,page]{appendix}

\include{aas_journals}
\usepackage{color}

\newcommand{\remove}[1]{}

\def\ie{{\frenchspacing\it i.e.}}

\def\be{\begin{equation}}
\def\ee{\end{equation}}
\def\ba{\begin{eqnarray}}
\def\ea{\end{eqnarray}}

\onecolumn

\begin{document}

\title{Astrophysical Tests of Modified Gravity: A Screening Map of the Nearby Universe}
\author{Anna Cabr\'e $^{1}$, Vinu Vikram$^{1}$,  Gong-Bo Zhao$^{2,3}$, Bhuvnesh Jain$^{1}$, Kazuya Koyama$^{2}$\\
 $^1$Center for Particle Cosmology, Department of Physics and Astronomy, University of Pennsylvania,
 Philadelphia, PA 19104\\
 $^2$Institute of Cosmology \& Gravitation, University of Portsmouth, Portsmouth, PO1 3FX, UK\\
$^3$National Astronomy Observatories, Chinese Academy of Science, Beijing, 100012, P.R.China}






\begin{abstract}
Astrophysical tests of modified modified gravity theories in the nearby universe have been emphasized recently by \cite{Hui:2009kc} and \cite{Jain:2011ji}. A key element of such tests is the screening mechanism whereby general relativity is restored in massive halos or high density environments like the Milky Way. In chameleon theories of gravity, including all $f(R)$ models, field dwarf galaxies may be unscreened and therefore feel an extra force, as opposed to screened galaxies. The first step to study differences between screened and unscreened galaxies is to create a 3D screening map. We use N-body simulations to test and calibrate simple approximations to determine the level of screening in galaxy catalogs.  Sources of systematic errors in the screening map due to observational inaccuracies are modeled and their contamination is estimated. We then apply our methods to create a map out to 200 Mpc in the Sloan Digital Sky Survey footprint using data from the Sloan survey and other sources. In two companion papers this map will be used to carry out new tests of gravity using distance indicators and the disks of dwarf galaxies. We also make our screening map publicly available. 
\end{abstract}

\section{Introduction}

Modified theories of gravity (MG) have received a lot of attention in
recent years, primarily as a way of explaining the  
observed accelerated expansion of the universe. 
A modification of general relativity (GR) on large (astrophysical) scales is generically 
expected to be a  scalar-tensor gravity theory, where a new scalar 
field couples to gravity. Equivalently
these can be described via a coupling of the scalar field  to 
matter which  leads to enhancements of the gravitational force. 
Nonrelativistic matter -- such as the stars, gas, and dust in galaxies --
will feel this enhanced force, which in general lead to larger dynamically
inferred masses.  The discrepancy from GR (or the true masses, measured 
via lensing) can be up to a factor of 1/3 in $f(R)$ 
gravity (for a review, see e.g.  \citet{Jain:2010ka} and \citet{Silvestri:2009}). 


The enhanced gravitational force should be detectable through  
fifth force  experiments, 
tests of the equivalence principle (if the scalar coupling to matter 
varied with the properties of matter) or through the 
orbits of planets around the Sun (Will 2006). 
\cite{Khoury:2003rn}
proposed that nonlinear screening of the 
scalar field, called chameleon screening, can suppress the fifth force 
in high density environments such as the Milky Way, so that Solar System and 
lab tests can then be satisfied. This screening was originally suggested 
to hide the effects of a quintessence-like scalar that forms the dark energy 
and may couple to matter (generically such a coupling is expected unless 
forbidden by a symmetry). Hence there are reasons to expect such a screening 
effect to operate in either a dark energy or modified gravity scenario. 
Qualitatively similar behaviour occurs in symmetron screening
(Hinterbichler \& Khoury 2010) and the environmentally dependent
dilaton (Brax et al. 2010). The screening map we  present here may be applied
 to these mechanisms as well once the parameters of the theories 
 are (approximately) matched. 

The logic of screening of the fifth force in scalar-tensor
gravity theories implies that signatures of modified gravity will
exist where gravity is weak. In particular,
dwarf galaxies in low-density environments may remain unscreened as the
Newtonian potential $\phi_{\rm N}$, which determines the level of screening,
is at least an order of magnitude smaller than in the Milky Way. Hence
dwarf galaxies can exhibit manifestations of modified forces in both
their infall motions and internal dynamics.
Hui, Nicolis and Stubbs (2009) 
discuss various observational effects, in particular the fact that 
rotation velocities of HI gas can be enhanced. Interestingly, 
stars can self-screen so that the stellar disk in an unscreened 
dwarf galaxy may have lower rotational velocity than the HI disk. 
Jain and VanderPlas (2011) study the offsets and warping of stellar disks, detectable by comparison of the morphology with the 
HI gas disk and through observations of 
rotation curves. 

Observations in the solar system rule out the presence of MG effects in galaxy halos with  masses higher than the Milky way, $M=10^{12}M_\odot$. In chameleon theories, two parameters are relevant for describing tests of gravity: the background field value $\chi_c$ and the strength of the coupling to matter $\alpha_c$ (using the notation 
of Jain, Vikram and Sakstein 2012). Models of $f(R)$ gravity have recently been 
worked out (Starobinsky 2007; Hu \& Sawicki 2007). In these models the parameter 
corresponding to the background field value is the present day value of the derivative of the function $f(R)$, denoted $f_{R0}$. Milky way constraints  
corresponds to  $f_{R0}>10^{-6}$. Cosmological probes rule out $f_{R0}>10^{-4}$  (based primarily on cluster counts, see Schmidt, Vikhlinin and Hu 2009 and Lombriser et al 2010; other cosmological tests are significantly weaker).  
Dwarf galaxies have the potential to probe field values down to $f_{R0}\sim 
10^{-8}-10^{-7}$.  Tests using distance indicators also probe field values 
down to $f_{R0}\sim 10^{-7}$. 

The tests with dwarf galaxies and nearby distance indicators rely on observations
of nearby galaxies, within 100s of Mpc and 10s of Mpc respectively. 
The first step in carrying out these tests is to classify the galaxies as
being screened or unscreened. The screened galaxies may be self-screened
if they are sufficiently massive, or screened by virtue of being close enough 
to a large galaxy or  group or cluster of galaxies. 

The goal of this paper is to provide a method to classify galaxies based on observables.  We also study systematics in the observations and its effect into screening. We use N-body simulations to test and calibrate our methods. We apply the methods to  galaxies in the SDSS.

\S 2 contains a description of the N-body simulations.  In \S 3 we describe the 
observable properties of galaxies used to quantify the level of screening. We test 
two different methods to make a screening map and explore a variety of systematic errors. We apply our results to the SDSS in \S 4 and conclude in \S 5.

\section{Description of simulations}

In order to find a proper criterion to determine whether or not a given galaxy is shielded against the modification of gravity, we utilise the high-resolution $N$-body simulations for the $f(R)$ gravity presented in \citet{Zhao:2010qy}. See also \citet{Li:2007, Li:2009, Li:2010, Li:2011}. The $f(R)$ gravity generalises the general relativity (GR) by adding to the Ricci scalar $R$ in the Einstein-Hilbert action  a specific function  $f(R)$, hence the name. If the function $f(R)$ is suitably chosen, this simple extension of GR can produce  accelerated expansion of the universe, and also evade the stringent solar system test thanks to the chameleon mechanism with which GR can be locally restored in  high density regions. Interestingly, this model predicted a different structure formation on the intermediate scales compared with the $\Lambda$CDM model, so it can be tested observationally (for a review of $f(R)$ gravity, see \citet{fR_review1} and \citet{fR_review2}).  $f(R)$ models may be considered representative of the chameleon mechanism, which is one of the viable screening mechanisms in modified gravity (MG). Others include the Vainshtein, symmetron and environmentally dependent dilaton screening.
    
We choose the $f(R)$ model proposed by \citet{Hu:2007nk}, and for the analysis we shall use the
high-resolution $N$-body simulations preformed by \citet{Zhao:2010qy} based on the Hu-Sawicki model. Specifically, 
\begin{equation} 
 f(R)=\alpha{R}/(\beta{R}+\gamma)
\end{equation}
 where $\alpha=-m^2{c_1},\beta={c_2},\gamma=-m^2,m^2=H_0^2\Omega_{\rm M}$
and $c_1,c_2$ are free parameters. The ratio $c_1/c_2$ determines the expansion rate of the universe in this model, and $c_1/c_2^2$, which is proportional
to $f_{R0}$ (the value of $|{\rm d}f/{\rm d}R|$ at $z=0$), dictates the structure formation. We follow \citet{Oyaizu:2008sr} to tune $c_1/c_2$ to match the expansion
history in a $\Lambda$CDM model, and choose values for
$f_{R0}$ so that those models can evade the
solar system tests. To satisfy these requirements, we set
$c_1/c_2=6\Omega_{\Lambda}/\Omega_{\rm M}$ and simulate four
models with $f_{R0}=10^{-4},10^{-5},10^{-6},10^{-7}$ ($F4$, $F5$, $F6$ and $F7$ hereafter). We do not use the $F4$ simulations for our screening tests. The models used and the corresponding Compton wavelength $\lambda_C$ are given in Table 1. The relation between $f_{R0}$ and $\lambda_C$ is approximately  $\lambda_C=32\sqrt{f_{R0}/10^{-4}}$ Mpc \citep{Hu:2007nk}.  

\begin{table}\label{tab:fro}
\begin{tabular}{ || p{2cm} | p{2cm} | p{2cm} ||}
	\hline
{\bf MODEL} &   {\bf $f_{R0}$}    &  {\bf $\lambda_C$ (Mpc)}     \\
	\hline
{\bf $F5$}   &  $10^{-5}$  &  $10$\\
	\hline
{\bf $F6$}  &  $10^{-6}$  &  $3$\\
	\hline
{\bf $F7$}  &  $10^{-7}$  &  $1$\\
	\hline
\end{tabular}
\caption{Table summarizing the models used in this paper. The background 
field value $f_{R0}$ and Compton wavelength $\lambda_C$ for the three model
simulations used are shown. }
\end{table}

The simulations were performed using a code based on MLAPM, which is a self-adaptive particle-mesh code. 
We used $256^3$ particles in a box 
of 64 Mpc/h a side, and preformed 10 realizations for each model to reduce the sample variance. The minimum particle mass is $10^{9}M_\odot/h$. In this paper we use the halo catalog, obtained by using the spherical over-density algorithm, with minimum mass set at $5\times 10^{11}M_\odot/h$ to obtain reliable dynamical masses. For each realization, 
all the models have the same initial condition at $z=49$, where GR is perfectly restored for these $f(R)$ models. We 
study in detail the $F6$ model since it gives us the best compromise of number screened/unscreened galaxies and 
this is the one that we study in more detail in data.



\begin{figure}[htb]
\begin{center}
\includegraphics[scale=0.28]{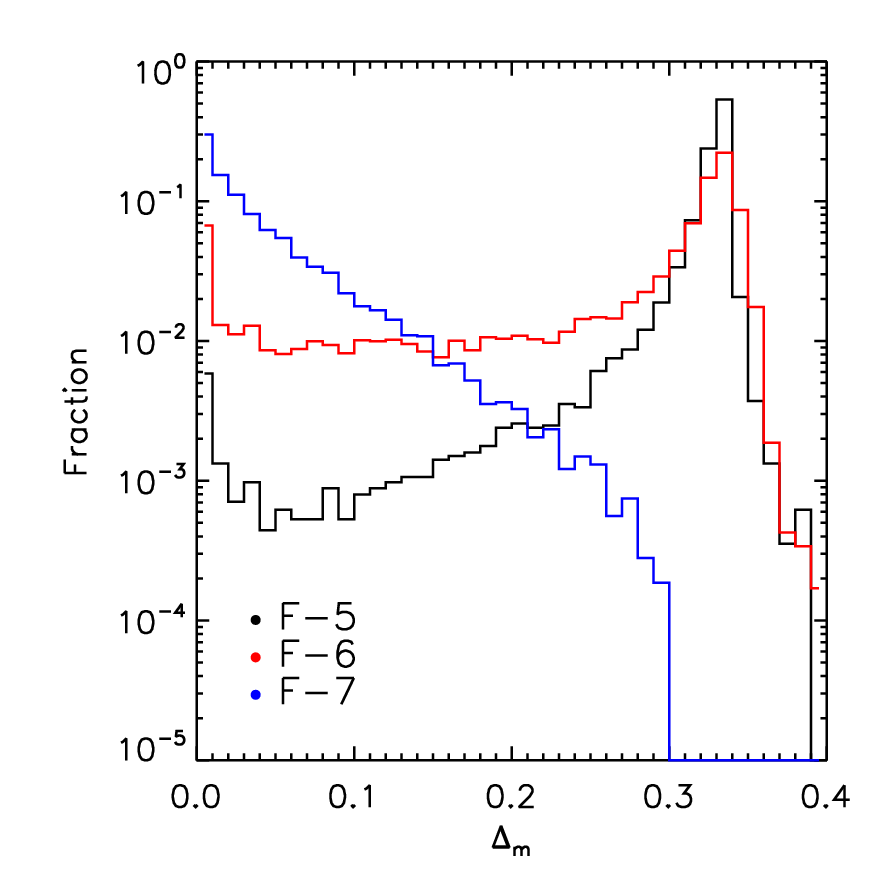}
\caption{Distribution of $\Delta_{\rm m}$ ($=M_D/M_L-1$) for different $f_{R0}$ models as labeled on the figure. A galaxy is less screened for large $\Delta_{\rm m}$. A fully unscreened galaxy has $\Delta_{\rm m}=0.33$ (higher values are due to simulation inaccuracies). As we move from the $F5$ to the $F7$ model, fewer halos are unscreened -- though a significant fraction still show small deviations, i.e. $\Delta_{\rm m}<0.1$.}
\label{fig:plotdeltam}
\end{center}
\end{figure}

\section{ Methods and Tests for Screening Criteria}

\subsection{Theoretical Background}

To test GR using dwarf galaxies, one of the first steps is to create a 3D screening map, which can be used to determine whether a galaxy is screened or not. In this section, using N-body simulation explained in the previous section, we develop a technique to find screened and unscreened galaxies. In the simulation we can estimate the degree of screening exactly by comparing the lensing mass and dynamical mass of the dark matter halos. We use $\Delta_{\rm m}$, the difference between the dynamical and lensing mass, defined in \cite{Zhao:2011cu}, to quantify the degree of screening for the simulated halos. However, in our initial application to observational   data we will use a binary classification, i.e. screened and unscreened. 

In the $f(R)$ gravity, the Poisson equation gets modified and reads, 
\begin{equation}\label{eq:poisson}
 \nabla^2\phi = 4\pi{G}a^2\delta\rho+\left(\frac{4\pi{G}a^2\delta\rho}{3}+\frac{a^2}{6}\delta{R}\right)=4\pi{G}a^2\delta\rho_{\rm eff}
\end{equation}
where $G$ is Newton's constant, $\phi$ is the gravitational potential (the time-time part in the perturbed FRW metric), $\delta{R}$ is the perturbed Ricci scalar due to MG, and $\delta\rho_{\rm eff}$ is
the perturbed total effective energy density, which contains
contributions from matter and modifications to the Einstein tensor
due to MG. The dynamical mass $M_D(r)$ of a halo is 
\be M_{D}\equiv\int{a^2}\delta\rho_{\rm eff}dV\ee where the integral
is over the extension of the body. Physically, the dynamical mass $M_D(r)$ means the mass
inferred from the gravitational
potential felt by a test particle at $r$. Assuming the
spherical symmetry, we can integrate the Poisson equation once to
obtain \be \label{eq:M_D} M_D(r)\propto{r^2}
d\phi(r)/dr. \ee  We actually use Eq. \ref{eq:M_D} to measure the dynamical masses of halos in our simulation. Note that if the gravity is modified, $M_D$ will be different from the lensing mass $M_{L}$, the true mass of an object since $M_D$ includes the contribution from
the scalar field, which mediates the finite-ranged fifth force within
the Compton wavelength. Therefore, the relative difference $\Delta_{\rm m}$, 
\be \Delta_{\rm m}\equiv
M_{D}/M_{L}-1\ee between the dynamical and lensing masses of the same object can be used as a indicator for MG. 

From Eq. \ref{eq:M_D}, we can estimate the range of $\Delta_{\rm m}$ by considering two extreme cases. \begin{description}
\item[(I)] In low density regions for unscreened galaxies, $\delta{R}$ can be dropped, giving $\Delta_{\rm m}=1/3$;
\item[(II)] In high density regions where GR is locally restored, namely, $\delta{R}=-8\pi{G}a^2\delta\rho$, then $\Delta_{\rm m}=0$.
\end{description} So $\Delta_{\rm m}\in[0,1/3]$, which can be used to quantify the extent of the screening. 

In Fig.  \ref{fig:plotdeltam} we show the degree of screening in different simulations. The figure shows the distribution of $\Delta_{\rm m}$ for simulations with different $f_{R0}$ values. The fraction of unscreened galaxies (galaxies with higher $\Delta_{\rm m}$) is largest for the $F5$ model -- the fraction decreases with smaller values of $f_{R0}$. Some values of $\Delta_{\rm m}$ are higher than 1/3, due to numerical uncertainties.

\subsection{Two Approximate Methods for Screening}
We are interested in a sample of galaxies within a few hundred Mpc, especially low mass galaxies.  Observations do not allow us to estimate lensing masses for these galaxies and thus estimate $\Delta_{\rm m}$ directly. This leads us to look for an alternate method to classify screened galaxies. We can estimate the dynamical masses of galaxies using measurements of luminosities or velocities. From the mass estimate we can find their virial radius under reasonable assumptions. Using the locations and 
dynamical masses of a reasonably complete galaxy catalog, we attempt to build a screening map using two simplified criteria as follows. 


In $f(R)$ gravity the chameleon effect recovers GR if $f_{R0} \leq 2/3\ |\phi_N|/c^2$, where $\phi_N$ is the Newtonian  potential of the object (Hu \& Sawicki 2007). This 
condition applies for an isolated spherical halo. We will make a simple approximation and use the same condition to determine whether an object is self-screened or environmentally screened. i.e. we look for the galaxies which satisfy the following conditions: 
\begin{equation}\label{eq:criterion} 
 \frac{|\phi_{\rm int}|}{c^2}>\frac{3}{2} f_{R0}\; \; \; \;\;\; \; \mathrm{or} \; \; \; \;\;\;\;\frac{|\phi_{\rm ext}|}{c^2}>\frac{3}{2} f_{R0}
\end{equation}
The first condition describes the  self-screening condition. However, low-mass galaxies may be \textit{environmentally} screened; the second condition is a simplified attempt at describing this,  with $|\phi_{\rm ext}|$ being the Newtonial potential due to the neighbor galaxies within the background Compton length $\lambda_C$. $|\phi_{\rm int}|$ for a galaxy is defined as  
\begin{equation}\label{eq:criterion21} 
 |\phi_{\rm int}|= \frac{GM}{r_{\rm vir}}
\end{equation}
where the mass $M$ and the virial radius $r_{\rm vir}$ are related via $\delta_{crit}=\frac{M}{4/3\pi r_{\rm vir}^3}$. Here we use $\delta_{crit}=200 \rho_{crit}$ where $\rho_{crit}$ is the critical density of the Universe. 
The external potential $|\phi_{\rm ext}|$ is evaluated using neighbors
\begin{equation}\label{eq:criterion22} 
 |\phi_{\rm ext}|= \sum_{d_i<\lambda_C+r_i} \frac{GM_i}{d_i}
\end{equation}
where $d_i$ is the 3D distance to the neighboring galaxy with mass $M_i$ and virial radius $r_i$, $\lambda_C$ is the Compton wavelength and $f_{R0}$ is the model parameter defined in \cite{Zhao:2010qy, Zhao:2011cu}. 

The  motivation for the above criteria is the following: the range of the fifth force is finite, defined by the Compton length $\lambda_C$. Our test galaxy will only feel the fifth force from neighbors within a distance $\lambda_C+r_i$ of the $i$-th galaxy (which itself must not be self-screened). We make the {\it ansatz} that the level of environmental screening is set by  $|\phi_{\rm ext}|$, though an accurate treatment requires a full solution of the scalar  field equations. As we increase $f_{R0}$, the fifth force range becomes larger. 

\begin{figure}[htb]
\includegraphics[scale=0.15]{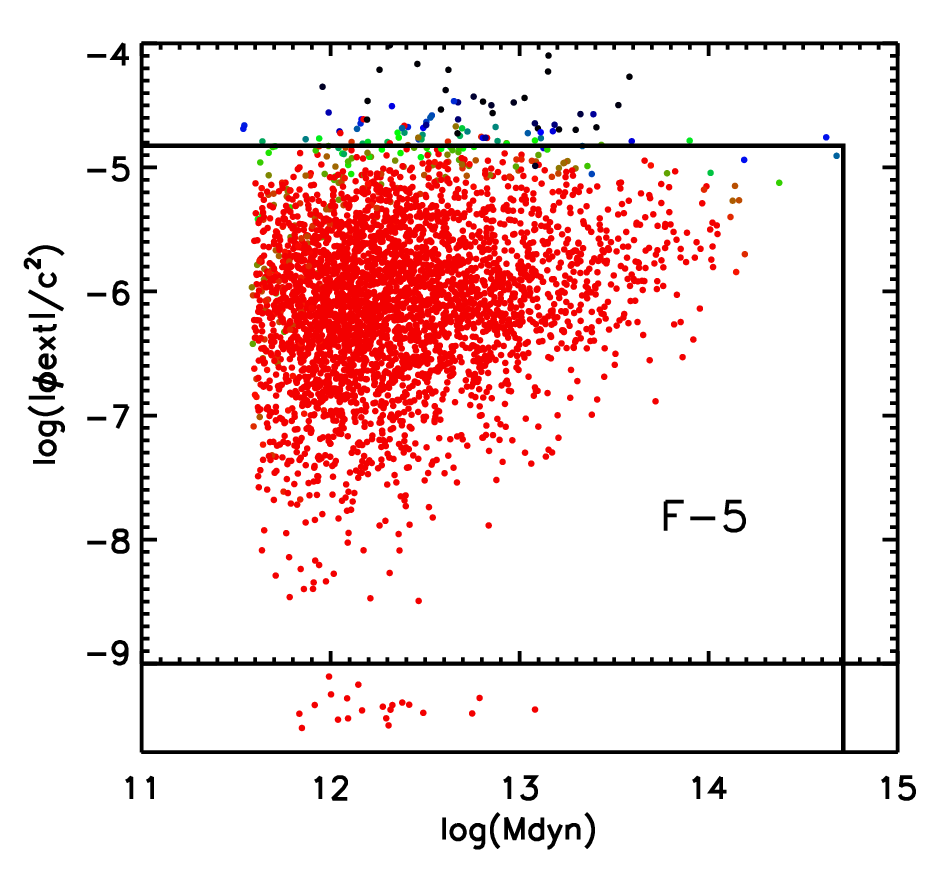}
\includegraphics[scale=0.15]{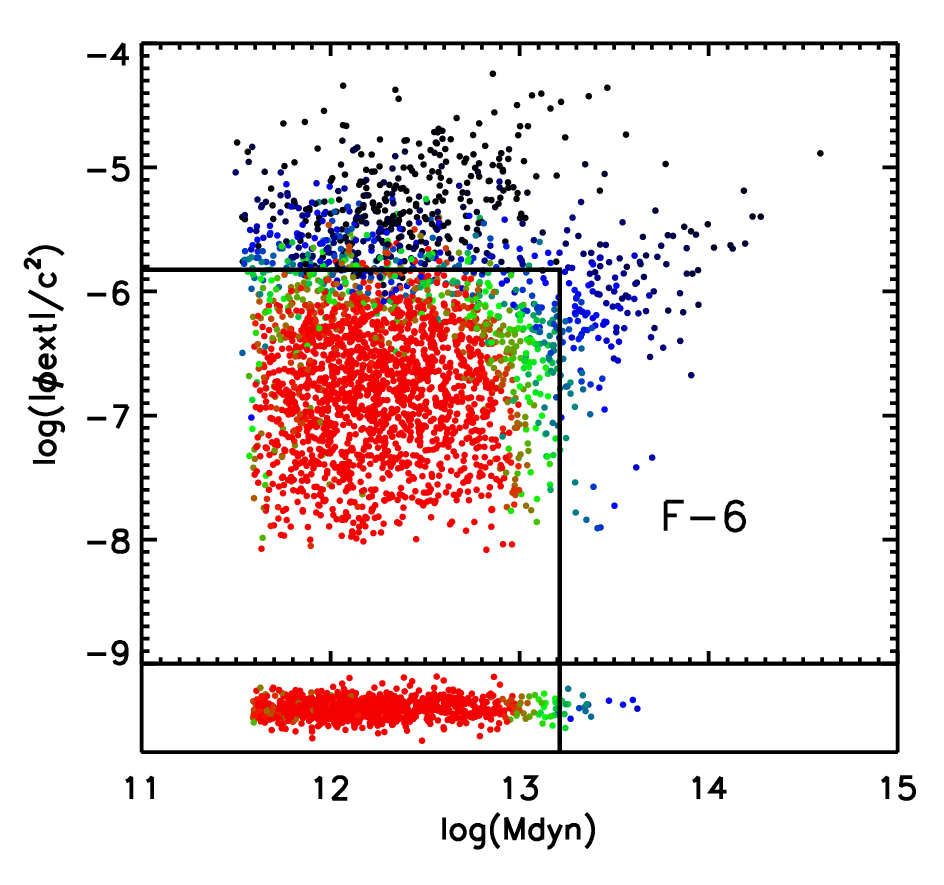}
\includegraphics[scale=0.15]{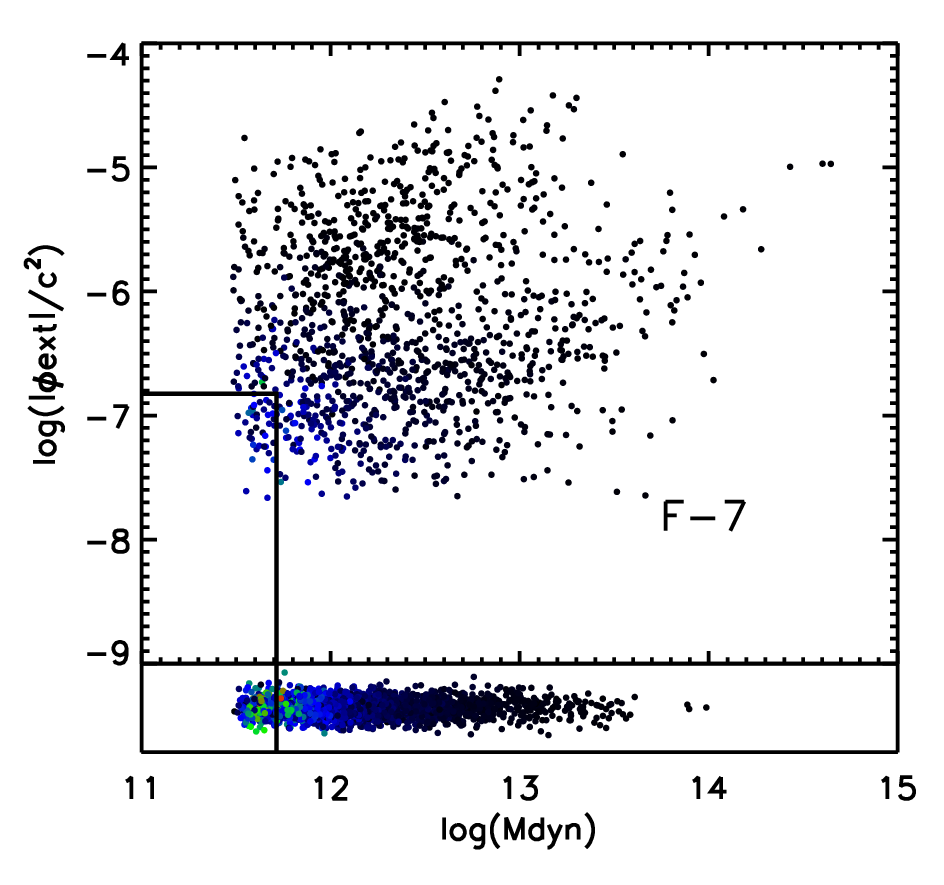}
\includegraphics[scale=0.15]{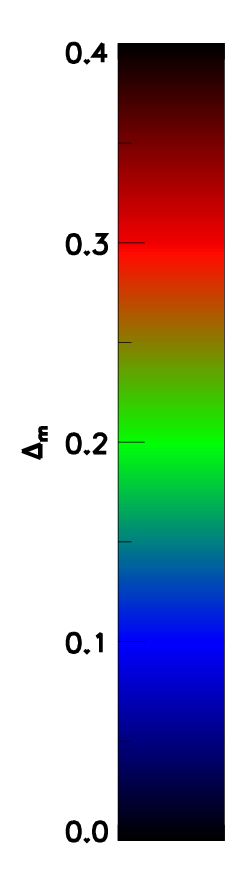}
\caption{Scatter plot for $|\phi_{\rm ext}|$ vs. $M_{\rm dyn}$ (equivalent to $|\phi_{\rm int}|$), colored according to $\Delta_{\rm m}$, as indicated in the colorbar on the right. Each point represents a galaxy halo from the N-body simulations. From left to right, we show different $f(R)$ models:  $F5$, $F6$ and $F7$. Blue colors indicate screened galaxies while red colors are unscreened. When $|\phi_{\rm ext}|=0$ we assign an artificial random value -- these halos are shown at the bottom of each panel. Note how the number of unscreened (red) galaxies is reduced when we move from $F5$ model to $F7$. The minimum non-zero $|\phi_{\rm ext}|$ is given by galaxies with one low-mass neighbor (limited by the simulation resolution) within a  Compton wavelength $\lambda_C$. A simple cut in $M_{\rm dyn}$ and  $|\phi_{\rm ext}|$ corresponding to  Eq. \ref{eq:criterion} seems to work to separate screened from unscreened galaxies.}
\label{fig:plotphimass}
\end{figure}

We use the N-body simulation described in the previous section to test the validity of the screening criteria of Eq. \ref{eq:criterion}. We use dynamical mass $M_{\rm dyn}$ from the simulation as it is the quantity that we can get easily from observations. The virial radius will also be obtained from $M_{\rm dyn}$. In Fig. \ref{fig:plotphimass} we show the screened and unscreened galaxies in the $|\phi_{\rm ext}|$--$M_{\rm dyn}$ plane. The points are colored according to the degree of screening ($\Delta_{\rm m}$). The figure shows that we can classify objects as screened (blue) and unscreened (red) using simply defined boundaries set by $f_{R0}$. The minimum non-zero $|\phi_{\rm ext}|$ is given by galaxies with just one low-mass neighbor (limited by the simulation resolution of $5\times 10^{11} M_\odot/h$) at a distance equal to   $\lambda_C$. For the $F7$ model  the self-screening condition is satisfied for nearly all resolved halos.  So we are not able to study the impact of environmental screening via the  external potential $|\phi_{\rm ext}|$ for this model. 

\begin{figure}[htb]
\includegraphics[scale=0.15]{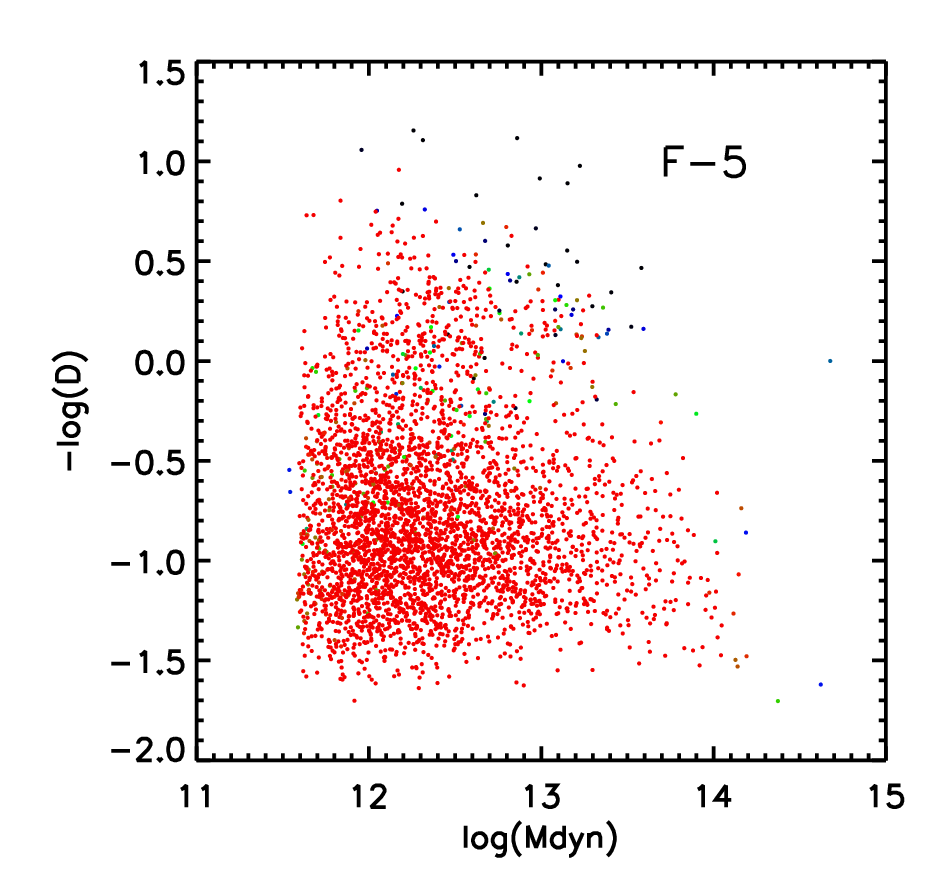}
\includegraphics[scale=0.15]{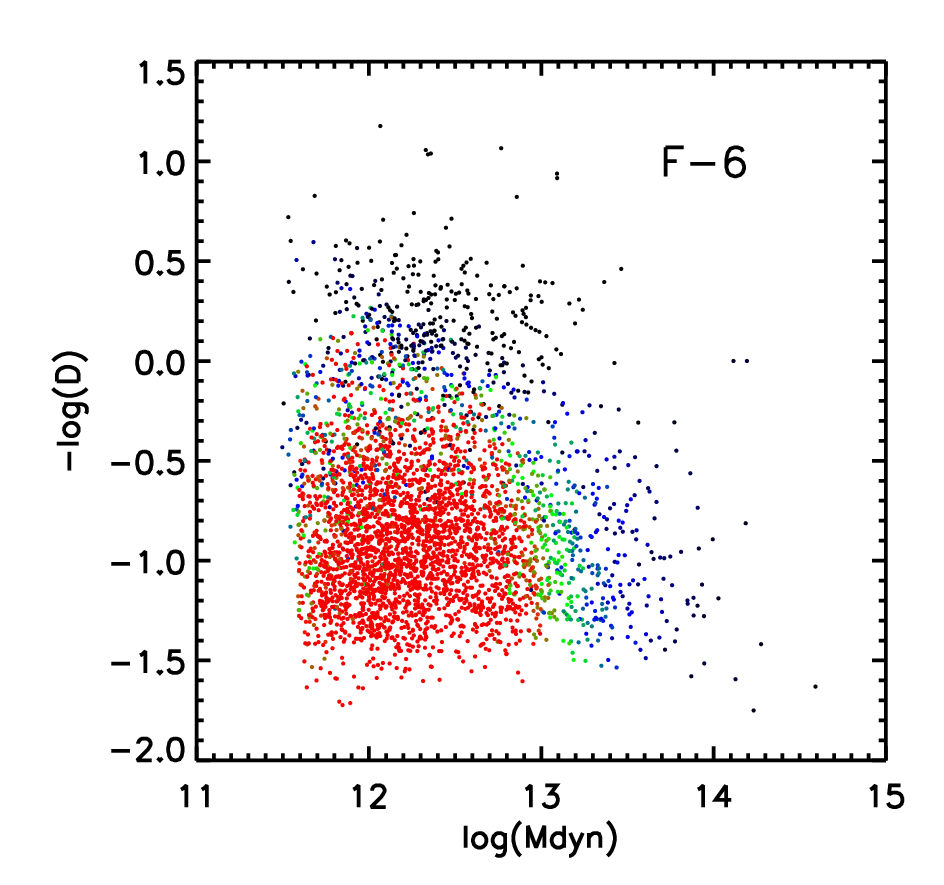}
\includegraphics[scale=0.15]{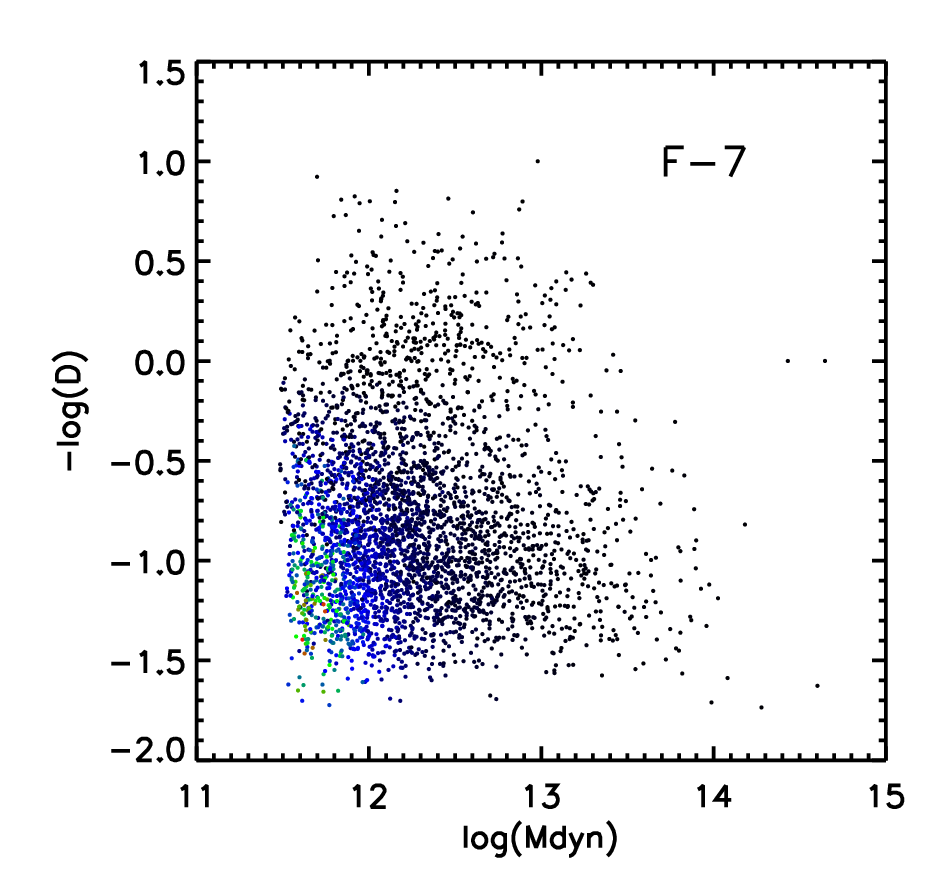}
\includegraphics[scale=0.15]{figures/colorbar-deltam.png}
\caption{Scatter plot for $D$ vs. $M_{\rm dyn}$, colored according to $\Delta_{\rm m}$, as shown in the colorbar on the right -- the three panels are as in Fig. \ref{fig:plotphimass}. 
 The distribution of galaxies in this plot is similar to Fig. \ref{fig:plotphimass}, so  a cut using $M_{\rm dyn}$ and $D$ is also likely to be effective in separating screened from unscreened galaxies. The effect of systematics is studied in subsequent figures. 
 }
\label{fig:plotDmass}
\end{figure}

As an alternate criterion for environmental screening  we study  the parameter $D_{fn}$ 
introduced by \cite{Haas:2011mt}.
For a test galaxy, it is defined in terms of its neighbors as: 
\begin{equation}
 D_{fn}=d/r_{\rm vir}
\end{equation}
where $d$ is the distance to the $n$-th nearest neighbor with mass $f$ times higher than the test galaxy, and $r_{\rm vir}$ the virial radius of the neighbor galaxy. Previous studies, \cite{Haas:2011mt}, show that $D_{11}$ ($D$, hereafter) is a good estimator for environmental screening. The values of $D$ are set by the threshold on $\Delta_{\rm m}$; for the $F6$ model, it is about 10 for a Milky Way sized galaxy with the threshold 
$\Delta_{\rm m}=0.1$. 
See Fig. \ref{fig:plotDmass} for the distribution of $\Delta_{\rm m}$ in the $D$ - $M_{\rm dyn}$ plane. Note how $|\phi_{\rm ext}|$ ($\sim M/d$) and $D$ ($\sim d / r_{\rm vir}$) are inversely correlated. However, the anti-correlation is not exact, since the weight in both estimators is different and $|\phi_{\rm ext}|$ uses information from multiple neighbors.  From Fig. \ref{fig:plotphimass} and Fig. \ref{fig:plotDmass}, we see that either $|\phi_{\rm ext}|$ or $D$ along with an estimate of $M_{\rm dyn}$ is a good proxy for screening. The combination of both estimators does not provide significant extra information. 
We find  below that the contamination is lower when using $|\phi_{\rm ext}|$.

However, this simple analysis may have limitations when one uses real data. In order to evaluate the performance for survey data, we  consider three cases below:
\begin{enumerate}
\item  The ideal case, \ie, without systematics. This allows us to study the intrinsic scatter around our cuts in $|\phi_{\rm ext}|$ or $D$ and $M_{\rm dyn}$.
\item Consider the contamination due to each type of systematics one by one.
\item  Consider the realistic case, \ie, all three  sources of uncertainties together.
\end{enumerate}

\subsection{Tests of Screening Without Systematic Errors}

Using simulations we can learn how to separate screened/unscreened galaxies using observables. We will describe screening using cuts in $|\phi_{\rm ext}|$, Eq. \ref{eq:criterion}, or $D$ and $M_{\rm dyn}$ (equivalent to $|\phi_{\rm int}|$). We have considered more sophisticated classification methods but the systematics in the observables limit us 
to simple approximations at this 
time. 
We focus our study on the $F6$ model, since the number of screened and unscreened galaxies is comparable there, and it is probably the most interesting case in terms of current limits. Towards the end, we will check how well our conclusions apply to other models.

\begin{figure}[htb]
\begin{center}
  \includegraphics[scale=0.25]{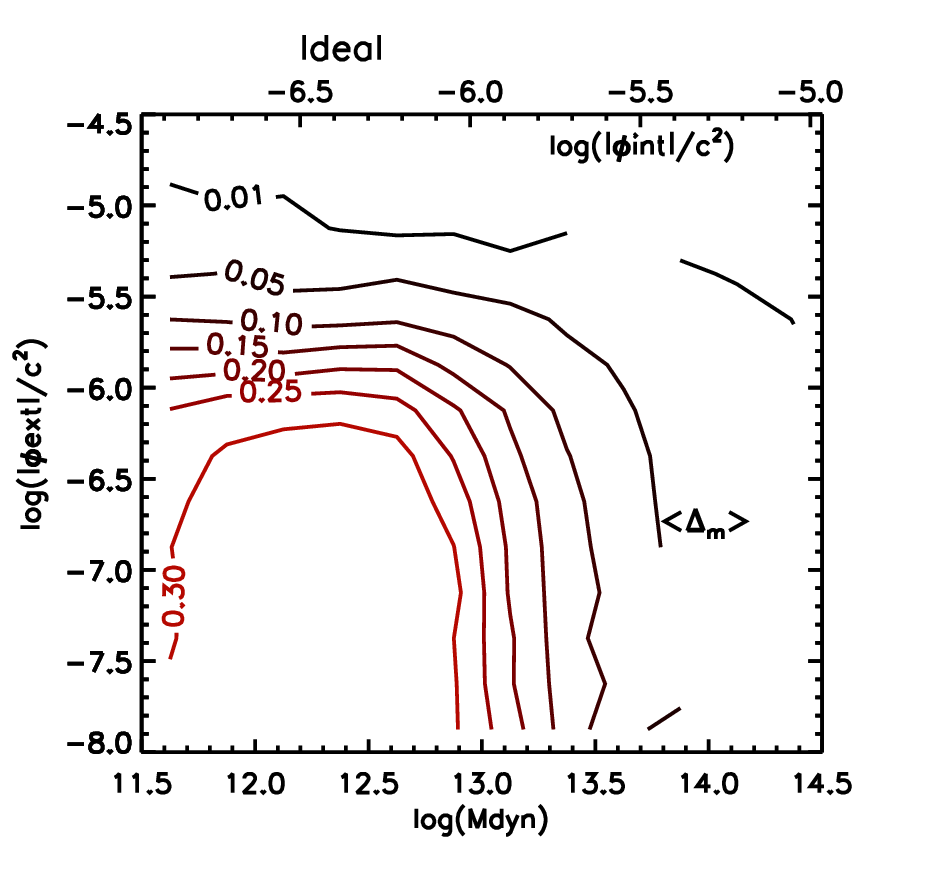}
\caption{Contours in mean $\Delta_{\rm m}$ for the ideal case of no systematic errors using
the $\phi_{\rm ext}$ method. 
$\Delta_{\rm m}$ increases from upper right to lower left in the plot. The contours generally follow  straight cuts in mass (or $|\phi_{\rm int}|$) and $|\phi_{\rm ext}|$. We have checked that even with systematics this plot remains  similar.} 
\label{fig:phimassmean}
\end{center}
\end{figure}

\begin{figure}[htb]
\begin{center}
  \includegraphics[scale=0.15]{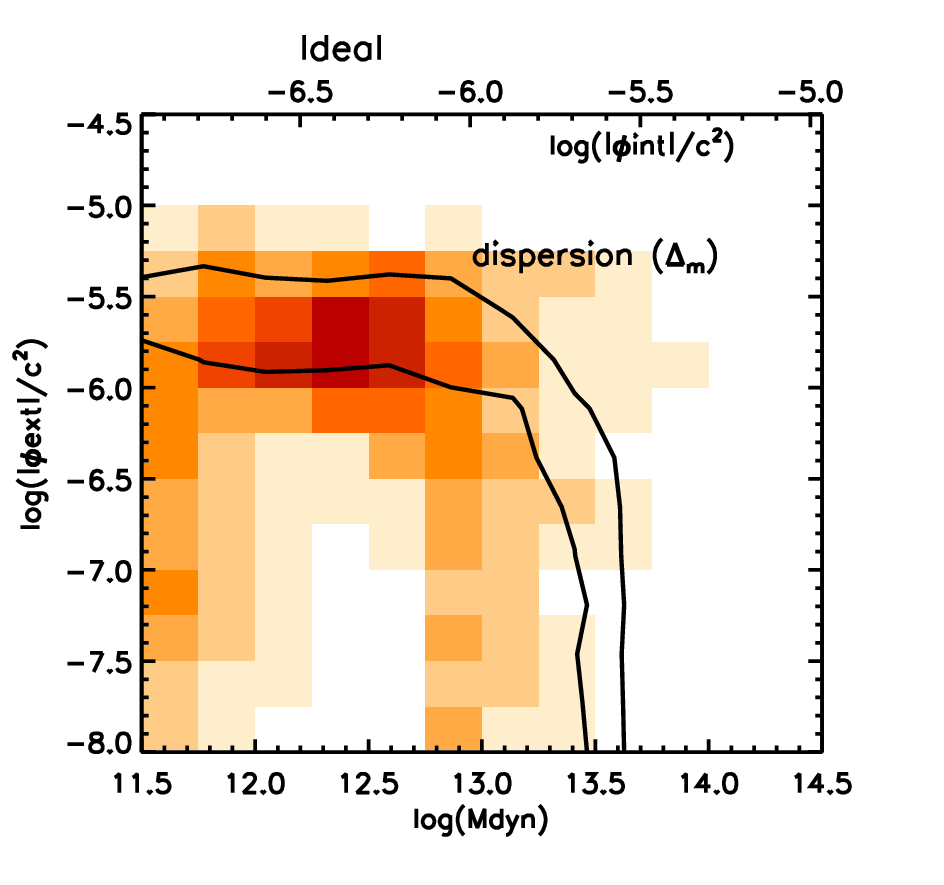}
  \includegraphics[scale=0.15]{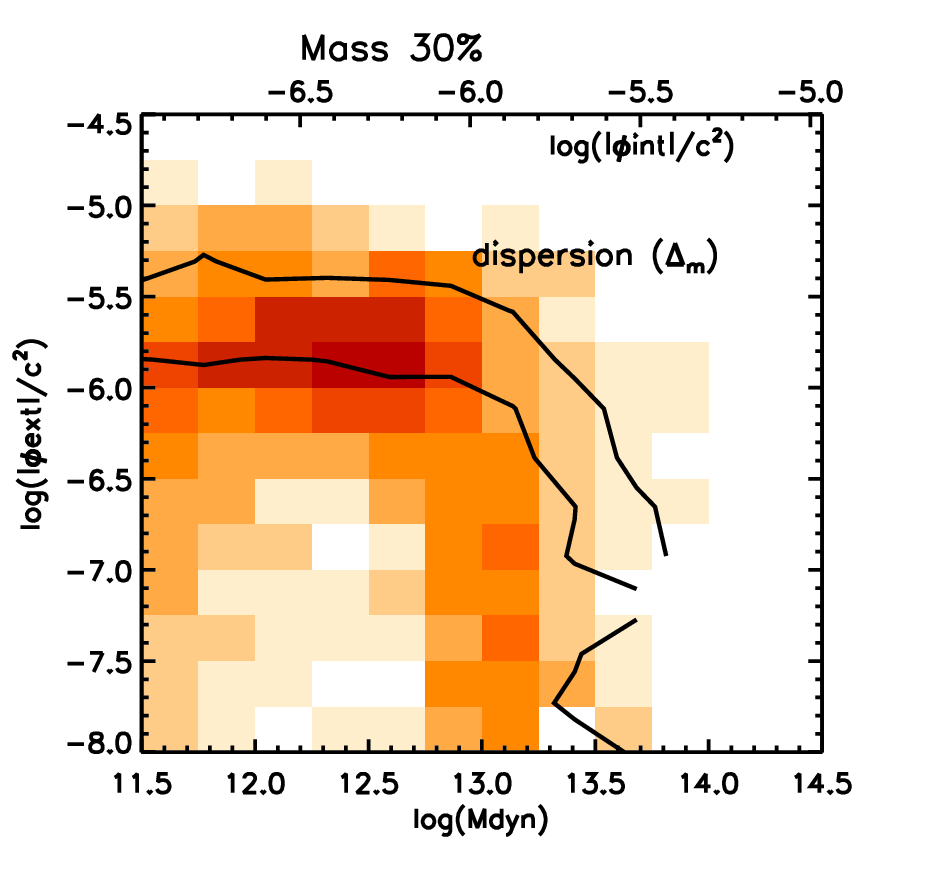}
  \includegraphics[scale=0.15]{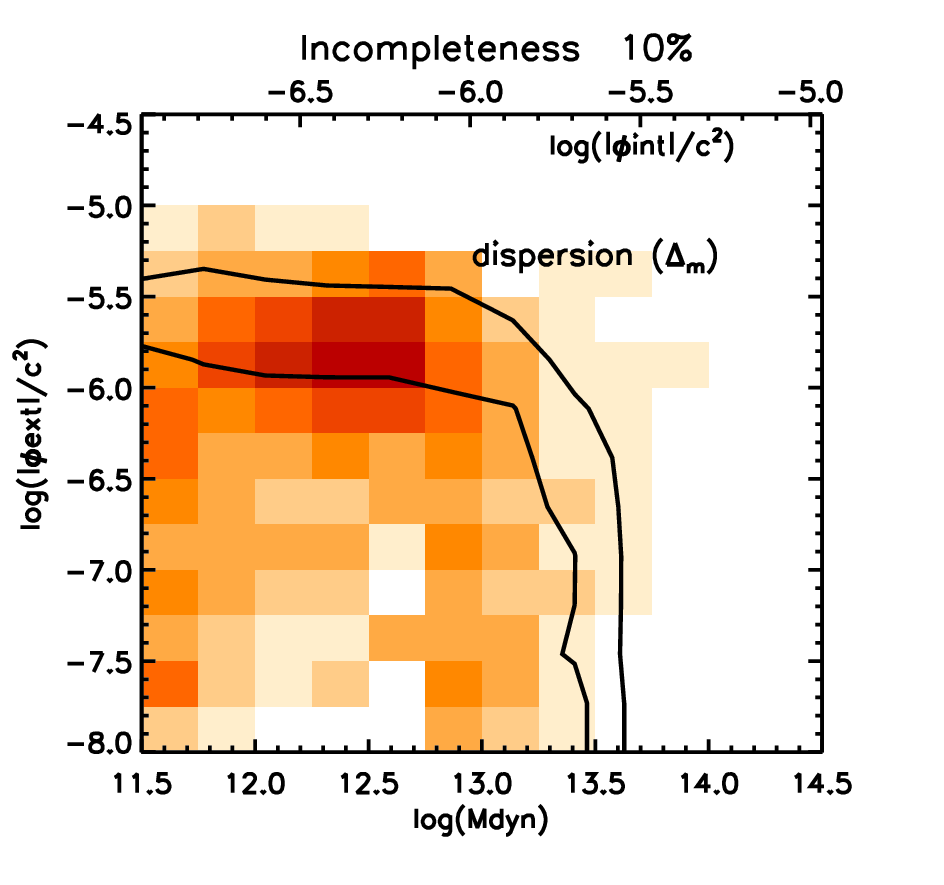}
  \includegraphics[scale=0.15]{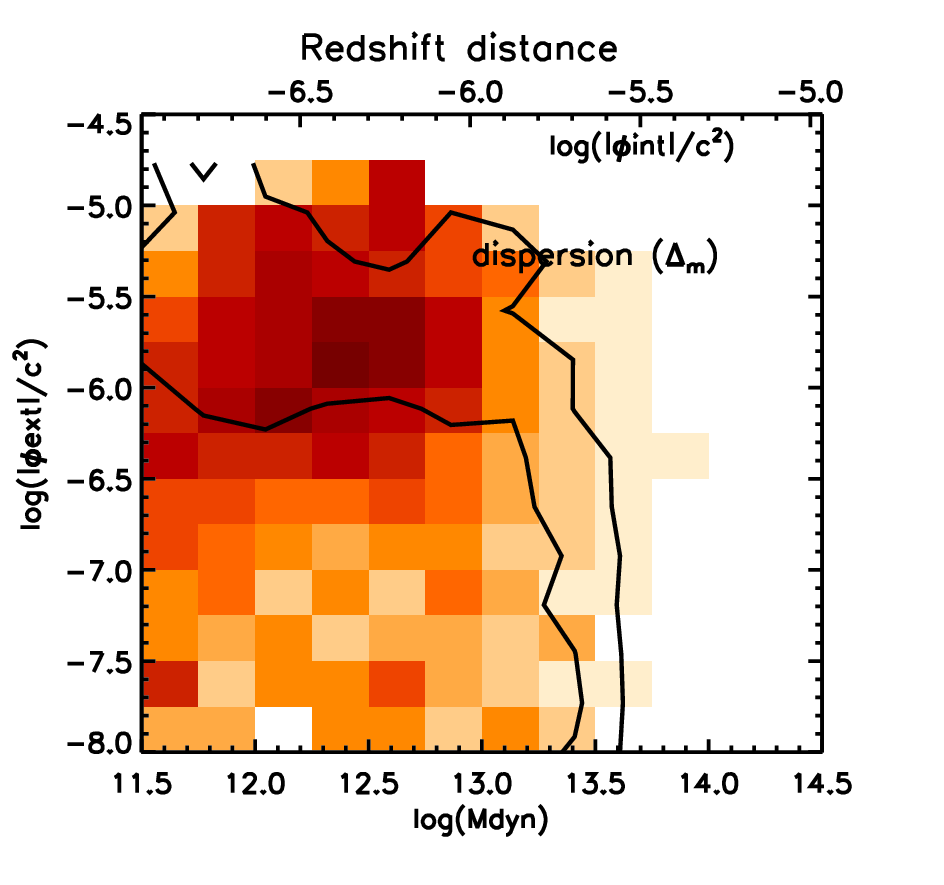}
  \includegraphics[scale=0.15]{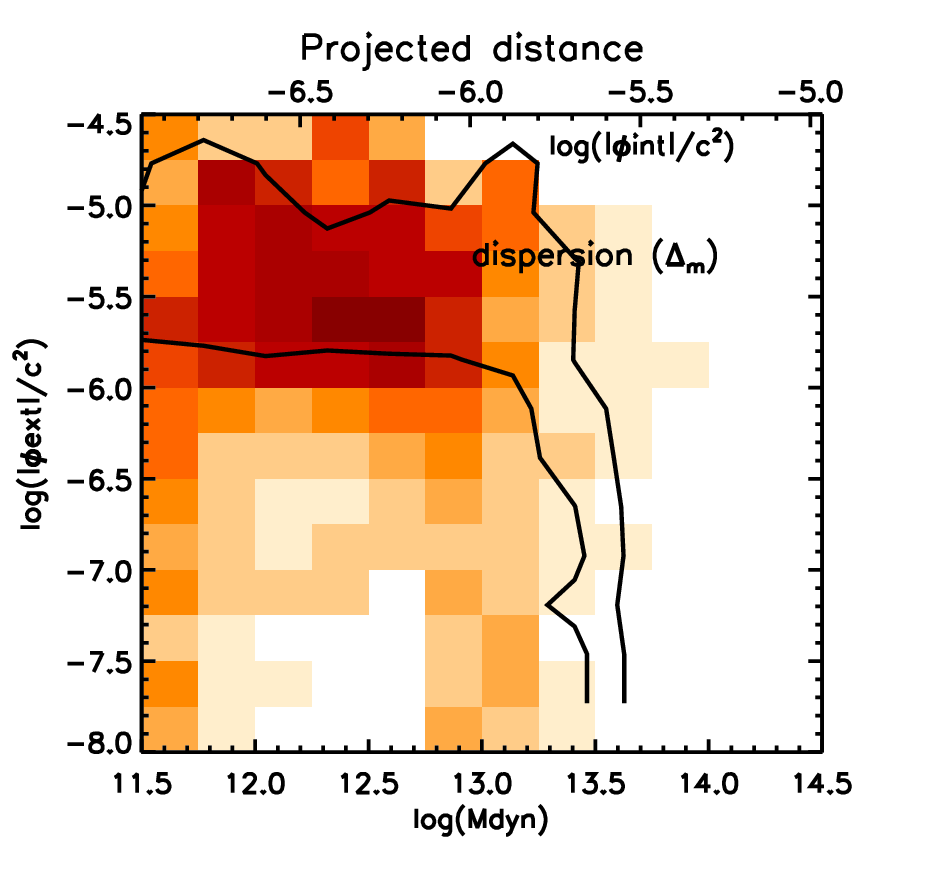}
  \includegraphics[scale=0.15]{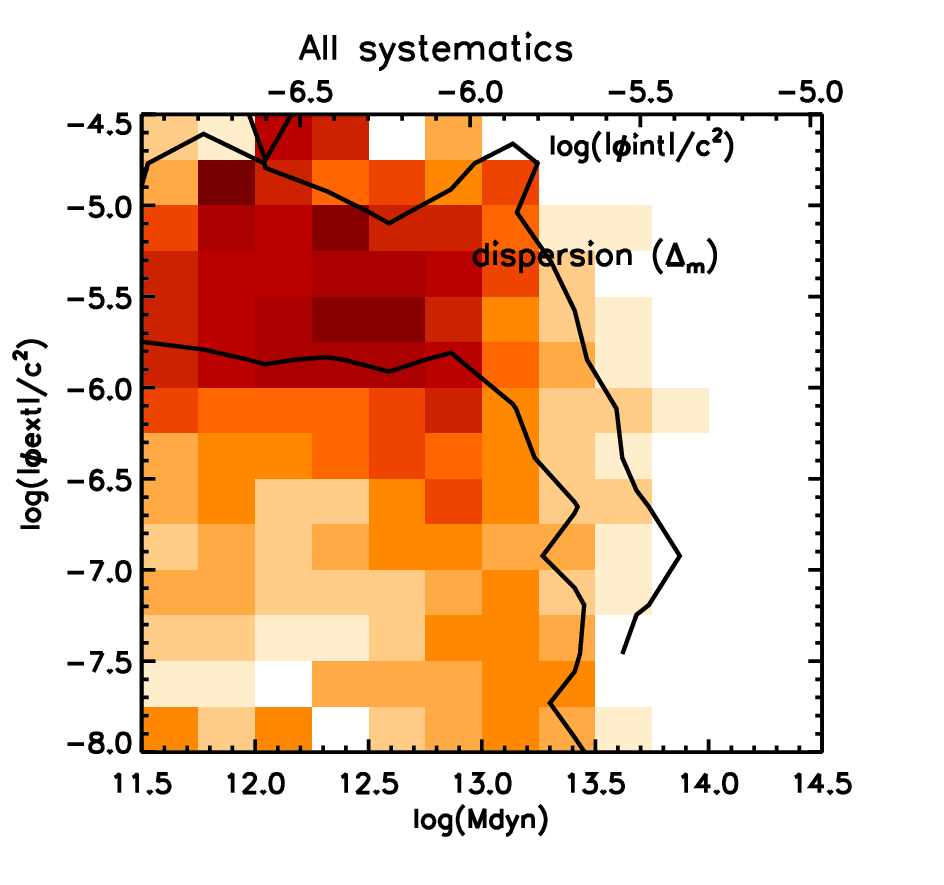}
\includegraphics[scale=0.14]{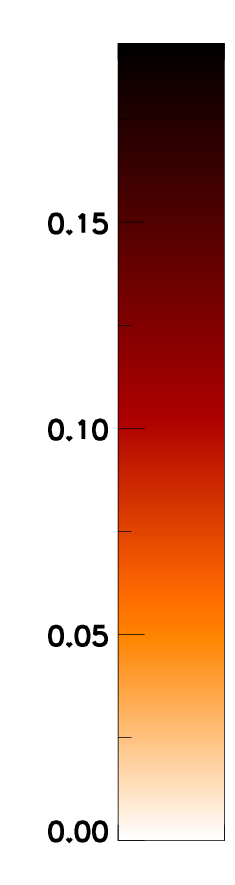}
\caption{The dispersion in $\Delta_{\rm m}$ is shown for the ideal case and for different systematics as labeled on the plots. The  colorbar on the right gives the scale. The ideal case (top-left) has some intrinsic dispersion, specially along the $|\phi_{\rm ext}|$ cut. We can see how the dispersion worsens once we add systematics -- top-center: mass uncertainty of 30\%; top-right: random incompleteness of 10\%; bottom-left: redshift distortion induced error in distance; bottom-center: use of projected distance and a line-of-sight cut of 500 km/s; 
bottom-right: all the previous systematics together (using projected distance).
 We overplot two solid lines that define the boundaries for 20\% (lower line) to 80\% (higher line)  of the screened galaxies per pixel (using the threshold $\Delta_{\rm m}=0.1$). }
\label{fig:phimassdisp}
\end{center}
\end{figure}

In Fig. \ref{fig:phimassmean} we show the contours of average $\Delta_{\rm m}$ in the $|\phi_{\rm ext}|$--$M_{\rm dyn}$ plane. We use bins of 0.25 in $\log(|\phi_{\rm ext}|)$ and $\log(M_{\rm dyn})$. We can see that $\Delta_{\rm m}$ increases towards low mass halos in low density environments (smaller $|\phi_{\rm ext}|$). We see in the figure that it is possible to make a division between screened and unscreened halos based on a fiducial value of $\Delta_{\rm m}$. It should also be noted that for any fiducial threshold value of $\Delta_{\rm m}$ the boundary looks mostly like a square defined by $M_{\rm dyn}$ on the x-axis and $|\phi_{\rm ext}|$  on the y-axis. In this paper we set $\Delta_{\rm m} = 0.1$ as the threshold value. Some part of the contamination evident in Fig. \ref{fig:plotphimass} comes from the dispersion of $\Delta_{\rm m}$ in the $|\phi_{\rm ext}|$--$M_{\rm dyn}$ plane. The top left panel of Fig. \ref{fig:phimassdisp} shows this dispersion for the ideal case. It can be seen that the dispersion is larger ($\sim 0.1$) near the boundary defined by the environmental potential, where the classification is not so clean. We study below how this dispersion changes with different systematics. The overplotted contour lines in Fig. \ref{fig:phimassdisp} demarcate 20\% (lower line) and 80\% (higher line) of screened galaxies (in each pixel).

At the end, the best cut will depend on the number of total galaxies. For the model $F6$, there are fewer screened galaxies than unscreened, hence the transition zone (where there is more contamination) will affect more the total contamination of screened galaxies.
We go one step further and calculate the best cuts using the F-score test, a common algorithm used in machine learning in order to classify objects. This algorithm takes into account that both samples might have different number of galaxies and also minimizes the contamination in both directions. This can be done by maximizing the F-score estimator, which is defined as: 
\begin{equation}
 F_{score}=\frac{2PR}{P+R}.  
\end{equation}
where $P$ is precision in screened galaxies and $R$ is recall, defined as follows. 
True positives (TP) are the screened galaxies that are tagged screened.
False positives (FP) are the unscreened galaxies that are tagged screened.
False negatives (FN) are the screened galaxies that are tagged unscreened. $P$ and 
$R$ are then defined as: 
$P=TP/(TP+FP)$ and $R=TP/(TP+FN)$.

\begin{figure}[htb]
\begin{center}
  \includegraphics[scale=0.15]{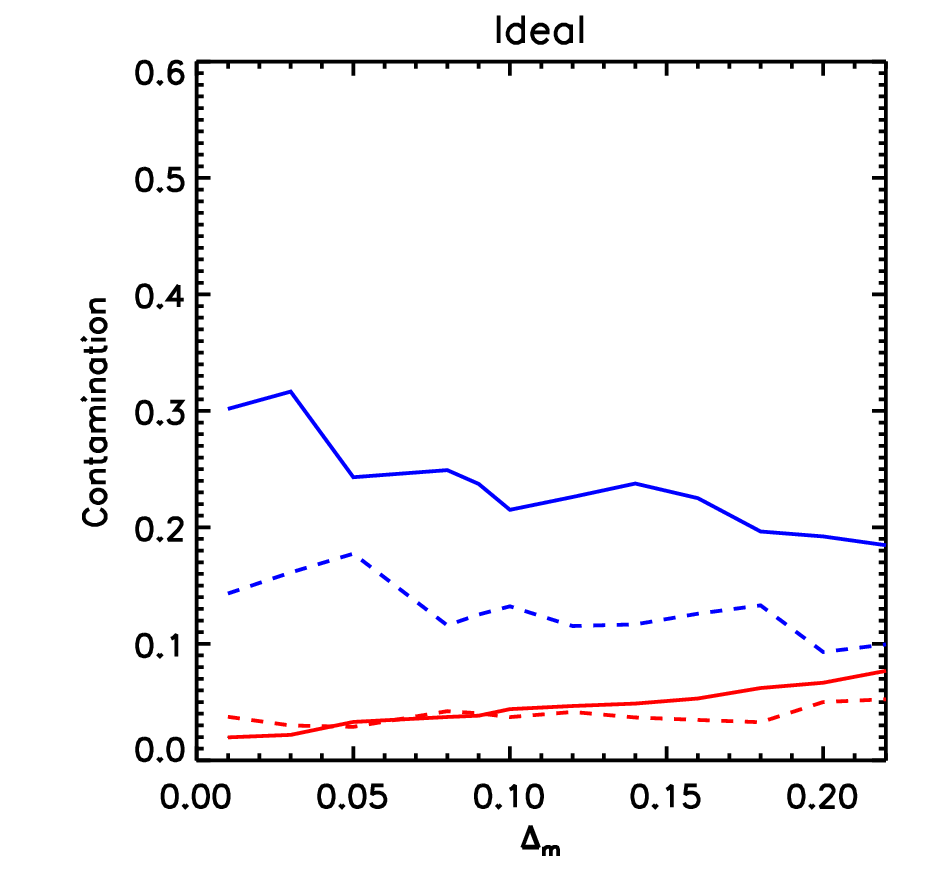}
  \includegraphics[scale=0.15]{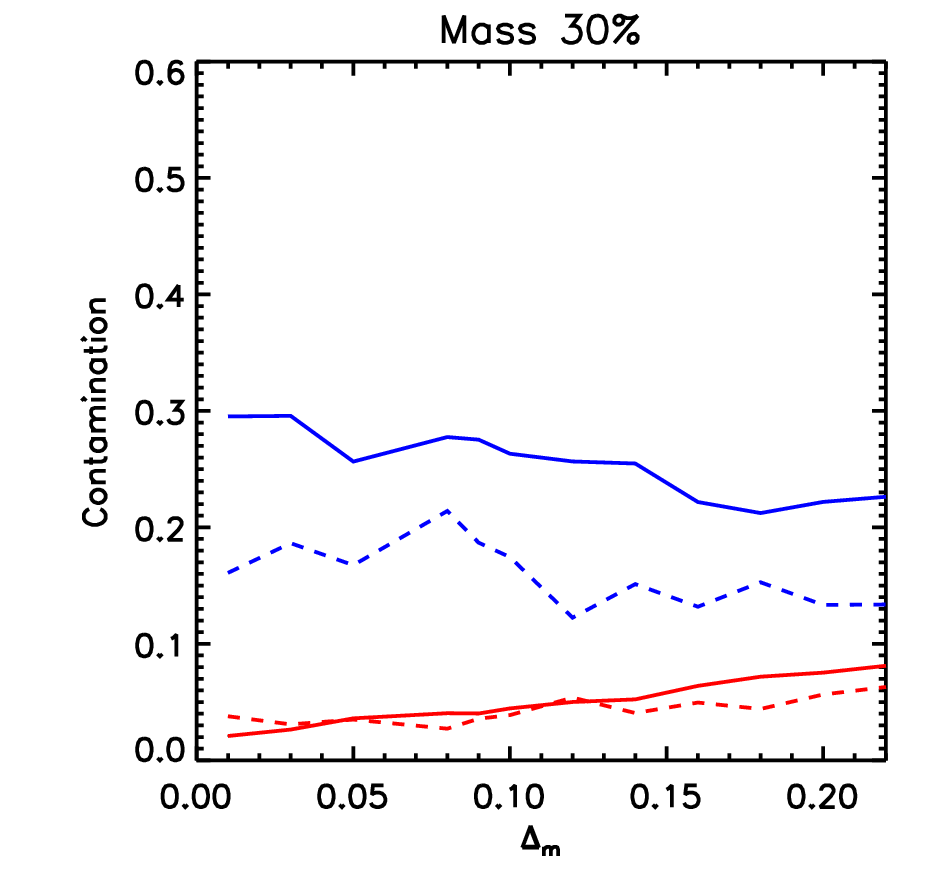}
  \includegraphics[scale=0.15]{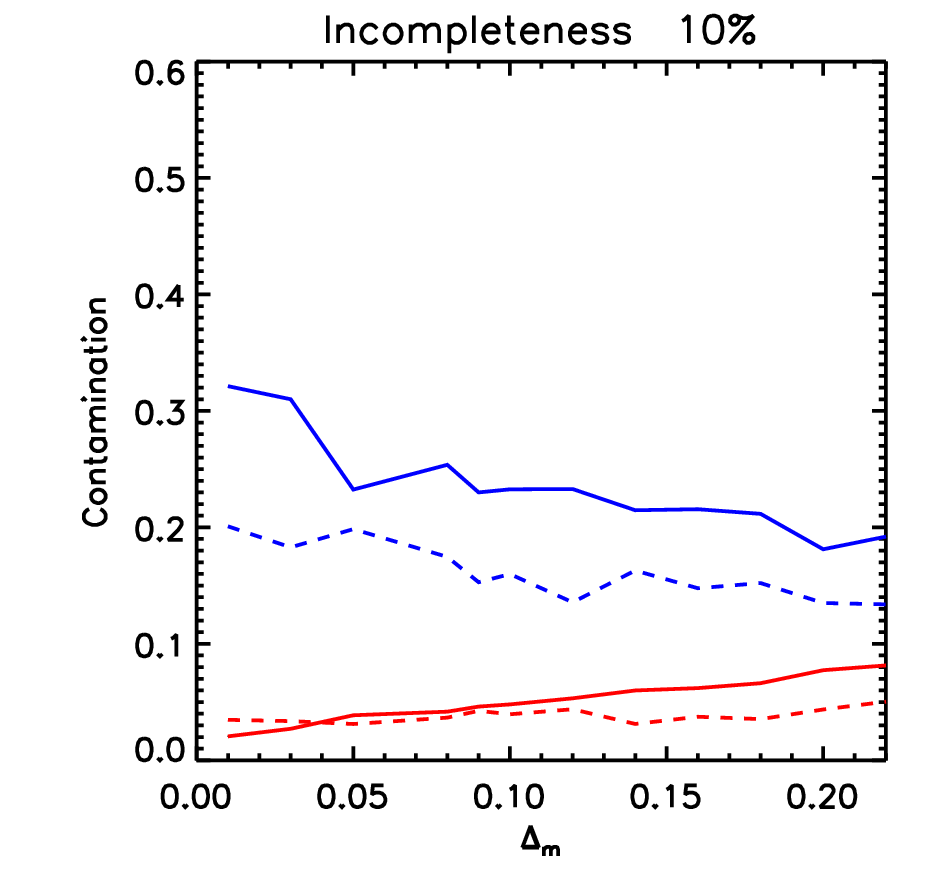}
  \includegraphics[scale=0.15]{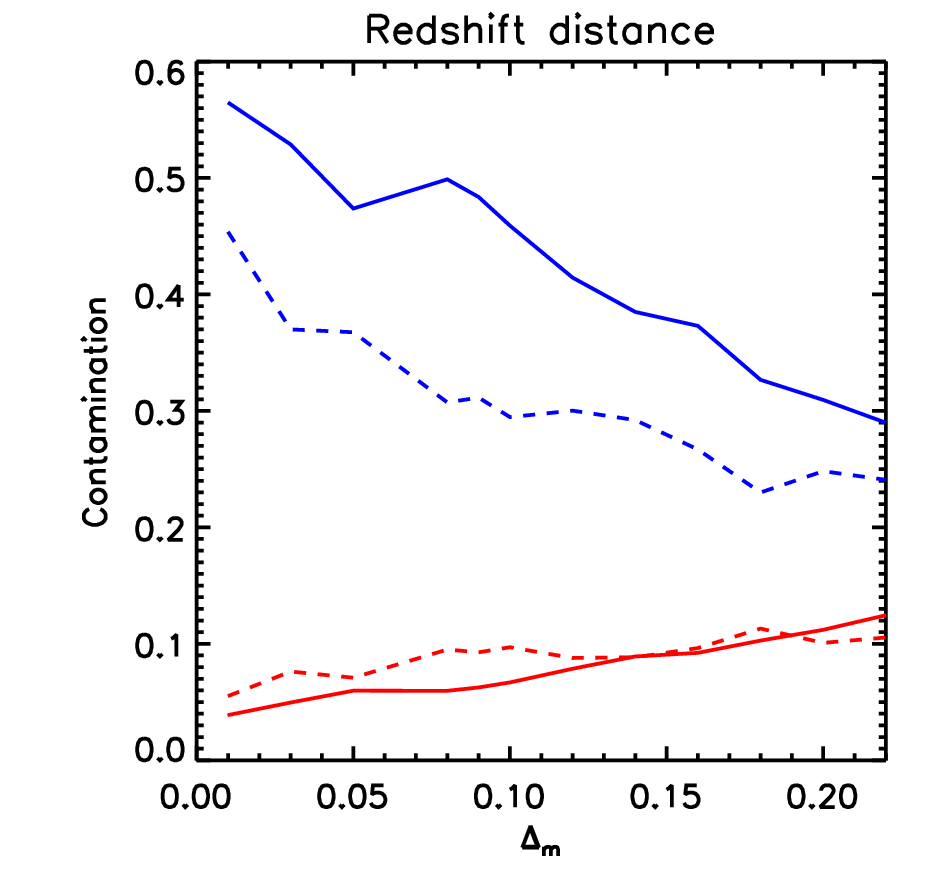}
  \includegraphics[scale=0.15]{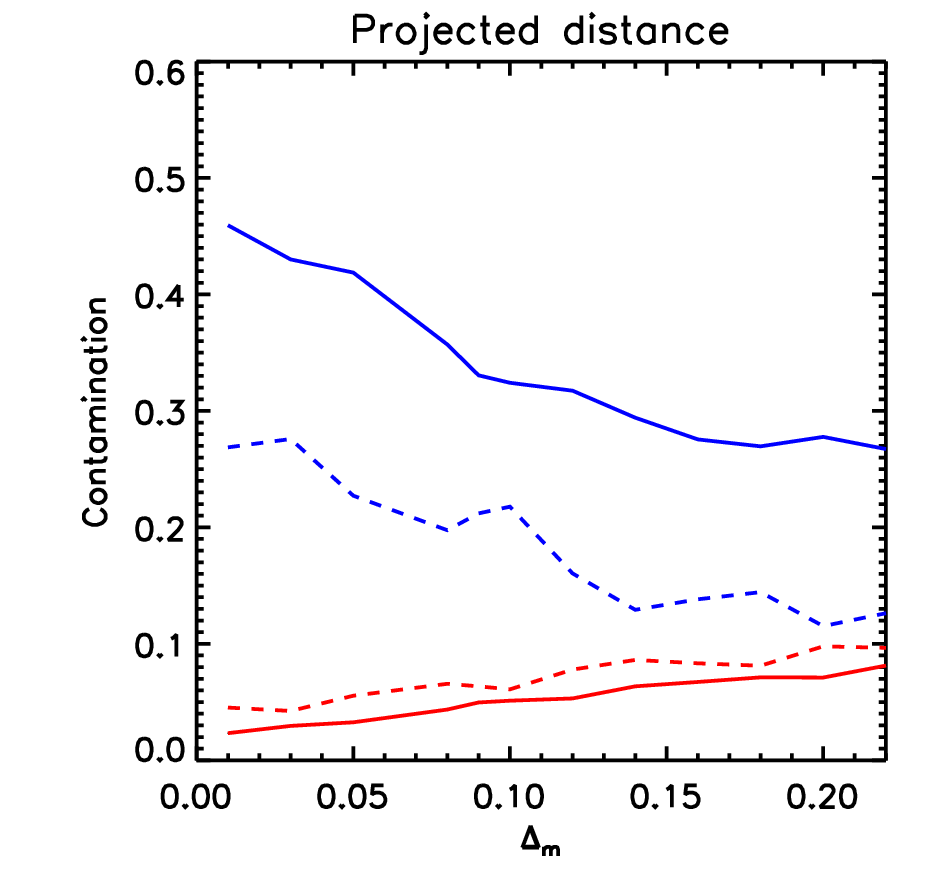}
  \includegraphics[scale=0.15]{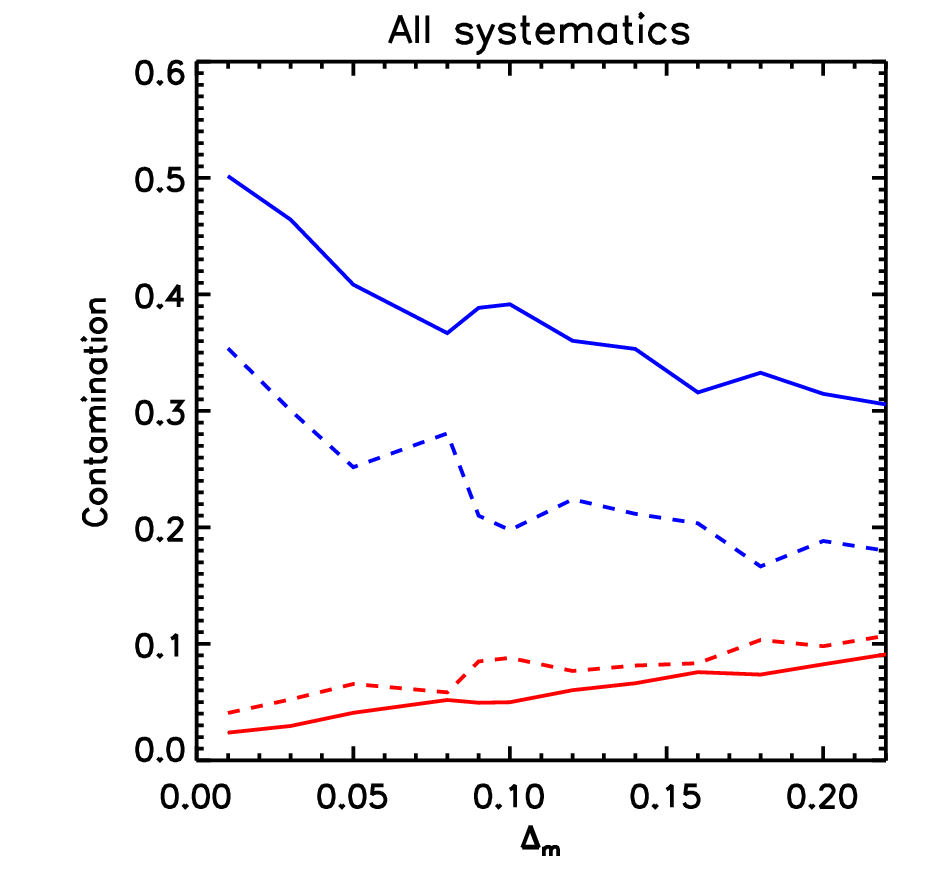}
\caption{The contamination in the galaxy sample using the  $|\phi_{\rm ext}|$--$M_{\rm dyn}$ cut (dashed lines) or $D$--$M_{\rm dyn}$ cut (solid lines).  The ideal case is shown in the top left panel and 
different systematics are shown in the rest of panels as labeled (see Fig. \ref{fig:phimassdisp} for description). We define the best cut as the one that maximizes the F-score for each $\Delta_{\rm m}$ that we use to separate screened from unscreened galaxies. The resulting contamination is plotted in blue for screened galaxies and red for unscreened galaxies. The fractional contamination in the unscreened sample is always lower because the  number of  unscreened galaxies is higher. Contamination in the screened sample when using $D$ is higher than when using $|\phi_{\rm ext}|$.}
\label{fig:plotcont}
\end{center}
\end{figure}

\begin{figure}[htb]
\begin{center}
  \includegraphics[scale=0.35]{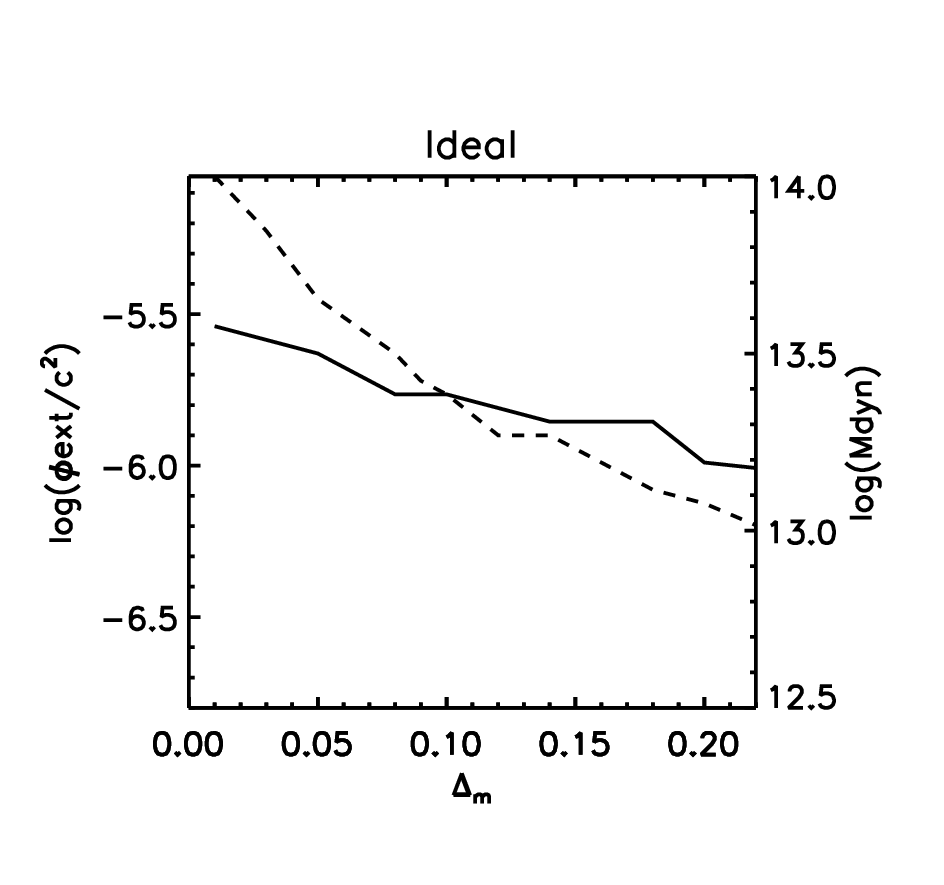}
\caption{We plot here the best cuts in $|\phi_{\rm ext}|$ (solid left axis) and $M_{\rm dyn}$ (dashed right axis) as a function of $\Delta_{\rm m}$ for the $F6$ model. When we change the threshold between screened and unscreened to higher $\Delta_{\rm m}$, we require a lower $|\phi_{\rm ext}|$ and lower $M_{\rm dyn}$ cut as we see in Fig. \ref{fig:phimassmean}. 
We have checked that the cuts are very similar even in the presence of the 
systematics tested in previous figures.}
\label{fig:plotcut}
\end{center}
\end{figure}

We plot the intrinsic contamination levels for screened (blue) and unscreened (red) galaxies for the $|\phi_{\rm ext}|$ - $M_{\rm dyn}$  method (dashed) or the $D$ - $M_{\rm dyn}$ method (solid) in Fig. \ref{fig:plotcont} (top-left). The first method gives lower contamination for screened galaxies (blue) in the range 10-15\%  for $\Delta_{\rm m}>0.1$. The contamination for unscreened galaxies is lower than 10\% always, and consistent for both methods. In Fig. \ref{fig:plotcut} we show the best cuts in $|\phi_{\rm ext}|$ (left axis,  solid line) and $M_{\rm dyn}$ (right axis, dashed line). The F-score is consistently around 0.8, which means that our suggested classification is robust.

\begin{figure}[htb]
\begin{center}
  \includegraphics[scale=0.2]{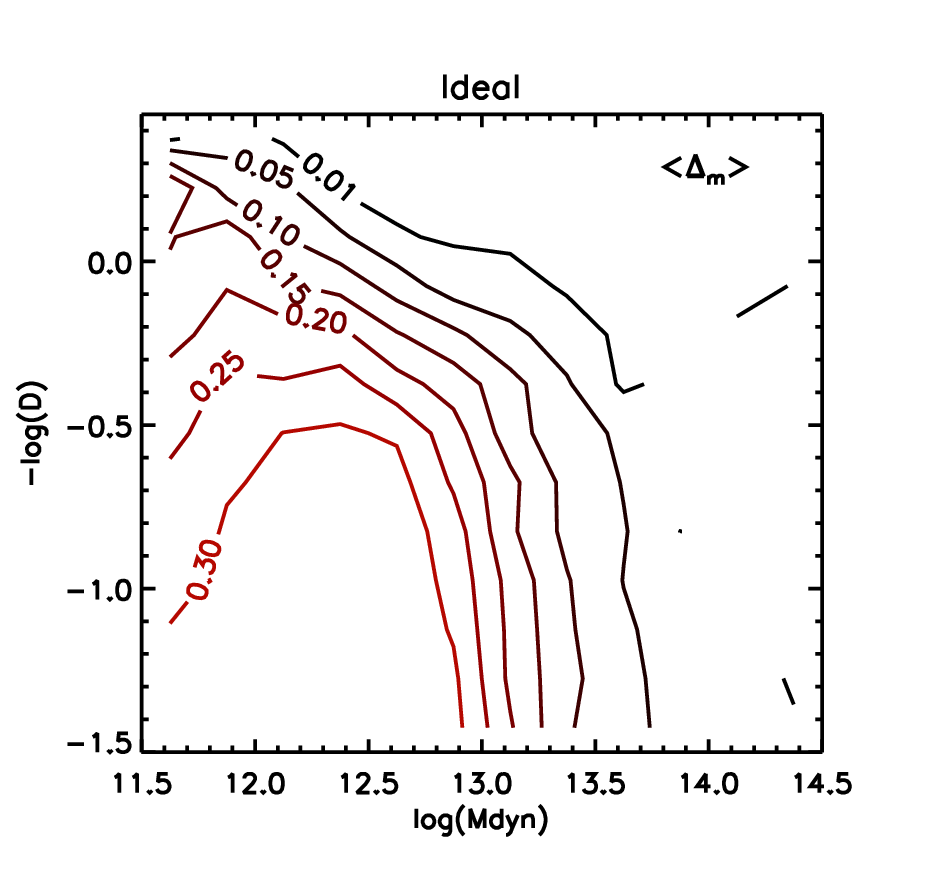}
  \includegraphics[scale=0.2]{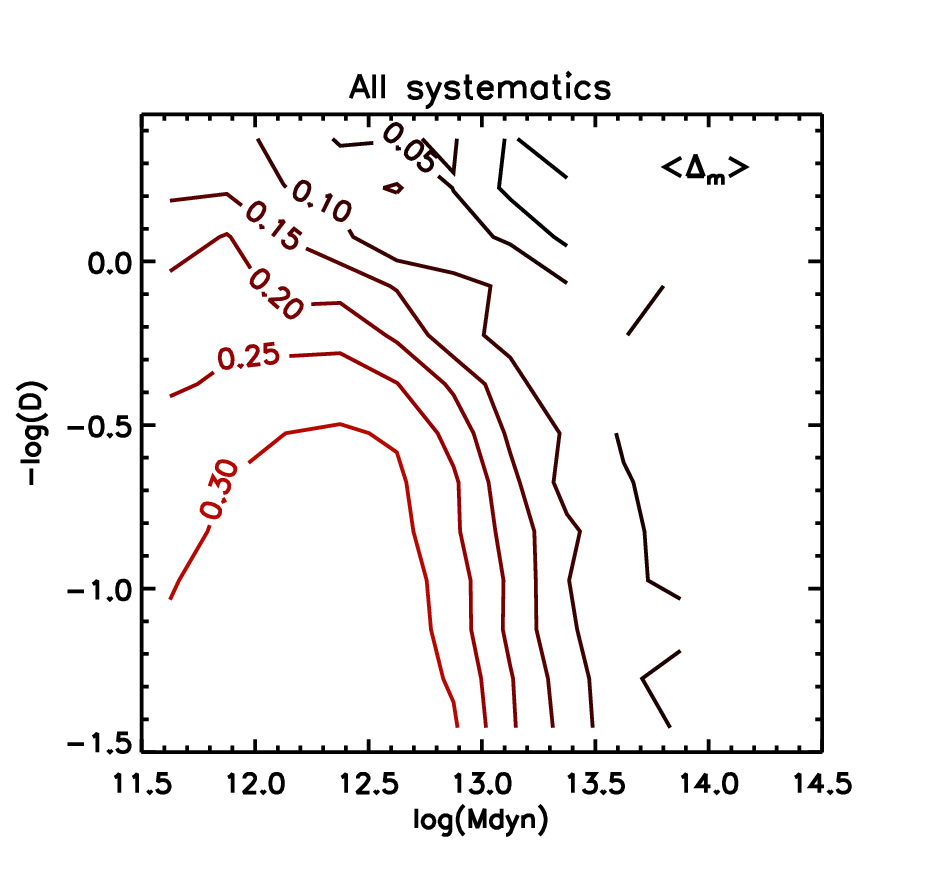}
  \includegraphics[scale=0.2]{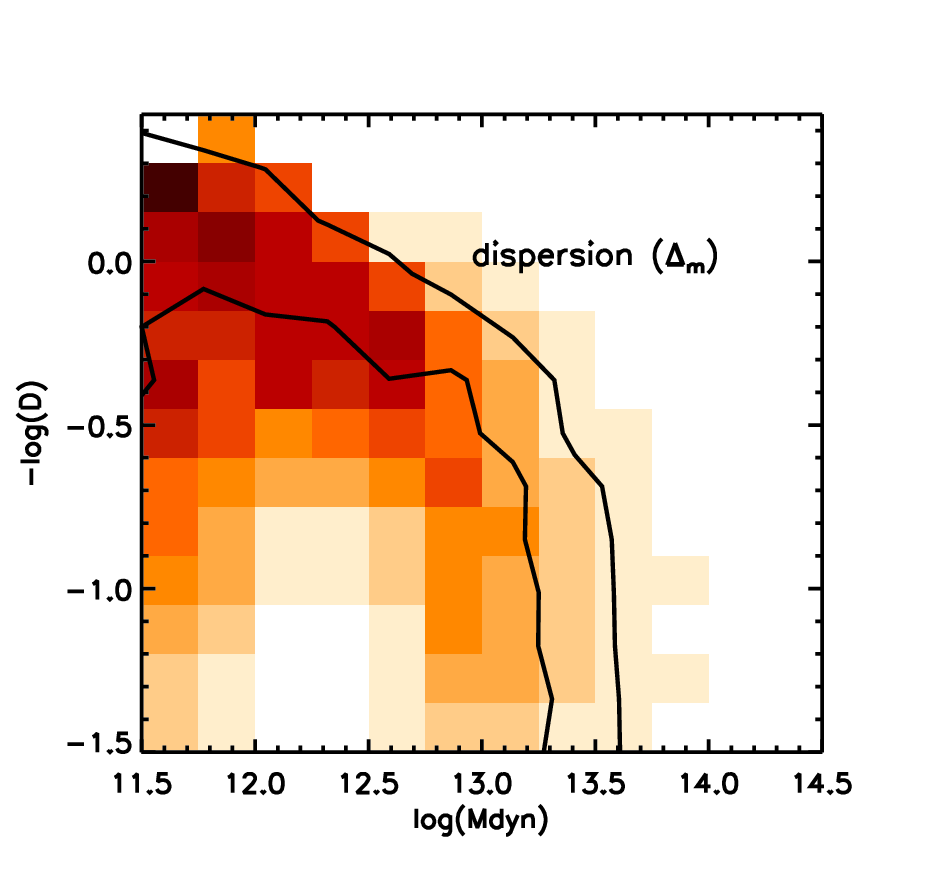}
  \includegraphics[scale=0.2]{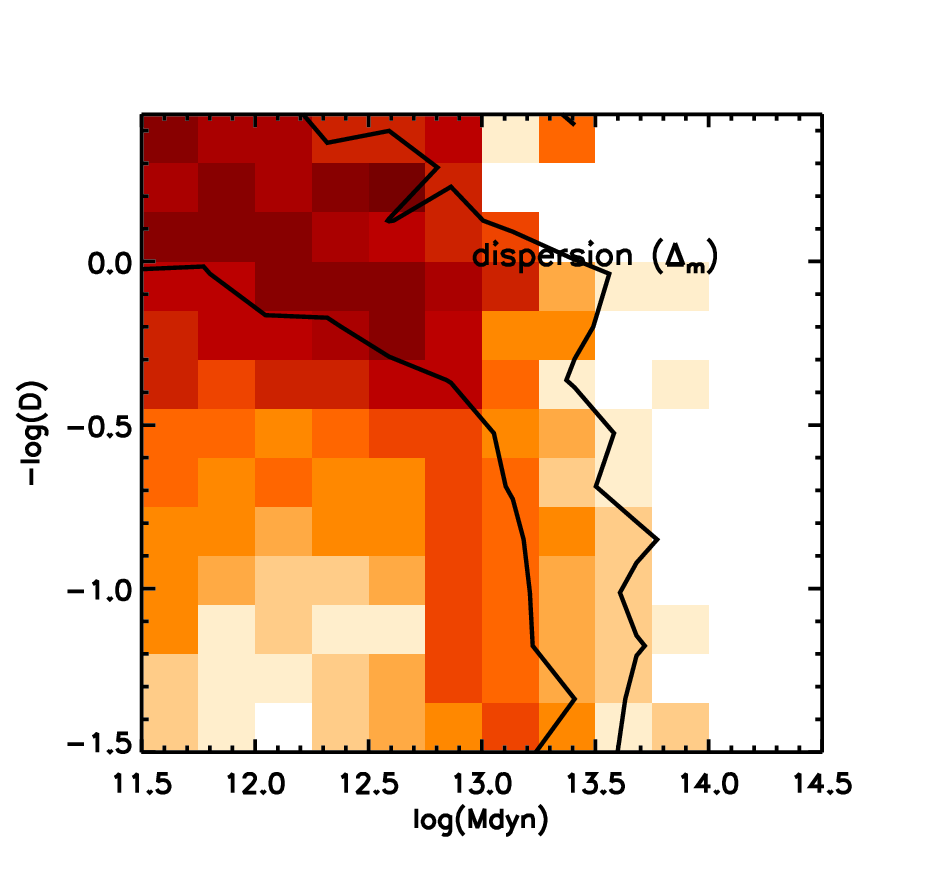}
\includegraphics[scale=0.2]{figures/colorbar-dispdeltamD.png}
\caption{Upper plots: Contours in mean $\Delta_{\rm m}$ for the ideal case (left)  and for all the systematics (right) for the environmental screening condition set by the neighbor distance $D$.  Note how the contour lines are not square-shaped, so a sharp cut in $D$ would not be accurate. Bottom: We plot the dispersion in $\Delta_{\rm m}$ for the ideal case (left) and all systematics together (right). Note that the dispersion is worse than in Fig. \ref{fig:phimassdisp}. We overplot two  solid lines as in the previous figure. }
\label{fig:Dmassmean}
\end{center}
\end{figure}

We have performed the same analysis using $D$ - $M_{\rm dyn}$ and we see in Fig. \ref{fig:Dmassmean}, top-left, that the cut is not as squared as in Fig. \ref{fig:phimassmean} so a sharp cut in $D$ and $M_{\rm dyn}$ would not be accurate. This is the main reason why a cut in $D$ - $M_{\rm dyn}$ does not work as good as $|\phi_{\rm ext}|$ - $M_{\rm dyn}$. In addition the dispersion, hence the contamination, is also higher. 
Note how contour plots in Fig. \ref{fig:phimassmean} give a cut in $|\phi_{\rm ext}|$ - $M_{\rm dyn}$ similar to the one obtained when using the F-score (Fig. \ref{fig:plotcut}), which is a good consistency check.

Now we extend our conclusions to other models. For $F5$, most of the galaxies are unscreened, only galaxies that reside near  big clusters ($M > 5\times 10^{14}M_\odot/h$) 
will be screened. We see in Fig. \ref{fig:plotphimass} that the proposed cut is valid. 
As for $F7$,  these simulations are not precise enough to study our proposed cut in detail, since we would need about one order of magnitude less massive galaxies to be confident of our tests. 

\subsection{Tests of Screening with Systematic Errors}

In this section we study how  systematic errors and other limitations in real surveys impact the  screening methodology described above.  There are three main sources of systematic errors in currently available data (in this paper we primarily use the  SDSS galaxy catalog). These are:
\begin{enumerate}
\item  Uncertainty in mass estimate. For the current  sample, 
the mass measurement is subject to 30\% uncertainty.
\item  Uncertainty in the distance estimate. The distance measurement is subject to 3-10 Mpc/$h$ (300-1000 km/s) uncertainty between galaxies in SDSS as the distances are derived from redshift measurements, which include peculiar velocities.
\item Incompleteness. Somewhere between 5-20\% of the neighboring galaxies for a typical test galaxy may be missing in our sample.
\end{enumerate}

In order to study the uncertainty in galaxy halo mass, we add a 30\% gaussian error to the dynamical mass in the simulations, both for test masses (the halo under classification) and neighbors. The virial radius is recalculated given the new estimated mass. See top-center plot in Fig. \ref{fig:phimassdisp} and Fig. \ref{fig:plotcont} to see the effect of mass uncertainty in the screening. The dispersion increases along the mass direction (Fig. \ref{fig:phimassdisp}). This implies that the contamination due to self-screening condition is larger than the contamination due to environmental screening condition. The best cuts in $|\phi_{\rm ext}|$ and $M_{\rm dyn}$ are similar to the ideal case. However, the contamination increases up to 5\% (for screened galaxies) with respect to the ideal case (Fig. \ref{fig:plotcont}).

Next, we study the completeness in the catalog. The incompleteness can arise in survey catalogues due to masking or by edge effects. To mimic these effects we randomly reduce the number of neighbors by 10\%. This incompleteness is independent of the mass. However, since there are more low mass galaxies in the universe, most of the randomly selected galaxies are  low-mass. The effects due to incompleteness can be seen on the top-right plots (Fig. \ref{fig:phimassdisp} and Fig. \ref{fig:plotcont}). In Fig. \ref{fig:phimassdisp}, we   see that this systematic affects mainly low massive galaxies. The contamination in screened galaxies increases by around 5\% (Fig. \ref{fig:plotcont}).

We also study the effect of uncertainties in distance on the classification. The distance uncertainty mainly originates from the use of redshift to estimate distances -- which is affected by the peculiar velocities of the galaxies. In order to introduce this systematic in the simulation, we add the peculiar velocity distortion $v_z$ to the $r_z$ direction as $r_z'=r_z+v_z/H(z)$. 
This systematic affects   $|\phi_{\rm ext}|$, since it changes the distance. It is a very large source of systematics as we can see in Fig. \ref{fig:phimassdisp}, bottom-left, and Fig. \ref{fig:plotcont}. We can reduce this systematic by estimating the distance using only the 2-D projected distance for suitably selected neighbors, and assuming spherical symmetry. 
We use neighbor galaxies within a line-of-sight distance of 5Mpc/h around the test galaxy (or equivalently 500km/s in the measured velocity). This is the velocity dispersion of an intermediate cluster. The improved results are on the bottom-center plot. 
We also checked the result using neighbors within a line-of-sight distance of 10 and 20 Mpc/h and found that the results remain the same. 
See bottom-left plot in Fig. \ref{fig:phimassdisp} for a summary of all the systematics together. 

We also tested what happens when there is incompleteness specifically in low mass neighbor galaxies used to determine $|\phi_{\rm ext}|$. Not surprisingly we find that 
completeness is important for halo masses just below the self-screening threshold. We tested this with the $F6$ simulations, and we expect that for the $F7$ model
this will be the principal limitation in our simulations. 

In conclusion, once we add systematics, the best cuts in $|\phi_{\rm ext}|$ and $M_{\rm dyn}$ remain nearly the same  as  in the ideal case. However, the overall contamination nearly doubles compared to the ideal case. 
We did the same analysis for the second classification scheme based on $D$-$M_{\rm dyn}$. The results are shown in 
Fig. \ref{fig:Dmassmean}. In the upper left panel we plot the contours of mean $\Delta_{\rm m}$ for the ideal case 
and in the upper right panel we show the contours with all systematics together. The bottom panels show the dispersion in 
$\Delta_{\rm m}$ for these two cases. We see that when adding the systematics the separator between screened and unscreened 
sample changes significantly from the ideal case. This makes these parameters harder to calibrate to reliably classify 
screened and unscreened galaxies. Also, the contamination is much larger than when using $\phi_{\rm ext}- M_{\rm dyn}$.
Thus in the presence of systematics the $|\phi_{\rm ext}|$--$M_{\rm dyn}$ is preferred.


\section{Screening Map for the SDSS Region}

Tests of gravity with a sample of galaxies requires a screening map: given the coordinates of a test galaxy, such a map would allow one to determine the level  
of environmental screening. 
We have created a screening map over about 10,000 square degrees out to a distance of 210 Mpc using the SDSS group catalog created by \cite{yang:2007} (DR7 version) and other surveys of galaxies and clusters. 

The group catalog from the SDSS covers the redshift range 0.01-0.2 and includes close to 400,000 galaxies. We restrict our screening map to $z=0.05$. 
Yang et al find group members using the friends-of-friends algorithm and calculate the group luminosity from 
member galaxies brighter than a certain threshold. 
The group luminosity is then converted to halo mass using an interative procedure that has been carefully tested with simulations. 
The mass of the groups ranges from $10^{11.6} M_\odot$ to $10^{15} M_\odot$. 
They estimated the scatter in the estimated mass using mock catalogs and found that it is less than $0.35$ dex. 

In the SDSS region, most of the contribution in our environmental screening map comes from the \cite{yang:2007} groups, but we include  other galaxy and cluster catalogs to determine $|\phi_{\rm ext}|$\footnote{NED provides a catalog of all the known extra galactic objects. We do
not use this  as many of the objects do not have absolute 
magnitudes which are required to estimate their masses. We do recommend  supplementing the catalogs described here with NED-based data on neighboring galaxies for tests  with a small enough sample of galaxies (so that case by case inspection is feasible). }. 
We use the group catalog from 2M++ all sky
survey \citep{lav011}. There are 3984 2M++ groups of galaxies in
our catalog. For local galaxies within 10 Mpc we use the galaxy catalog of   \citet{kar04}. 
These two catalogs improve our completeness for lower mass halos but not uniformly with distance. 

We compiled a catalog of clusters with known redshifts using archival data. Our
catalog mainly consists of Abell  \citep{abe89} and ROSAT \citep{ebe96} bright
clusters. We ignore many clusters from the Abell catalog as they lack spectroscopic
redshifts or lie beyond $z=0.1$. Our  catalog has a total of 675
clusters of galaxies. 

A direct estimate of the masses of clusters and groups of
galaxies can be found either by X-ray observation or by lensing observations.
However, many of our clusters and groups do not have such an estimate. So we
used  established scaling relations to estimate the mass of
clusters and groups. We use either the mass-luminosity relation given by
\citet{rei02} or the mass-velocity dispersion relation given by \citet{evr08} to
determine the mass. For the galaxies within 10 Mpc we use the mass estimates given by \citet{kar04}, which covers  313 out of 451 galaxies.

\begin{figure}[htb]
\begin{center}
  \includegraphics[scale=0.47]{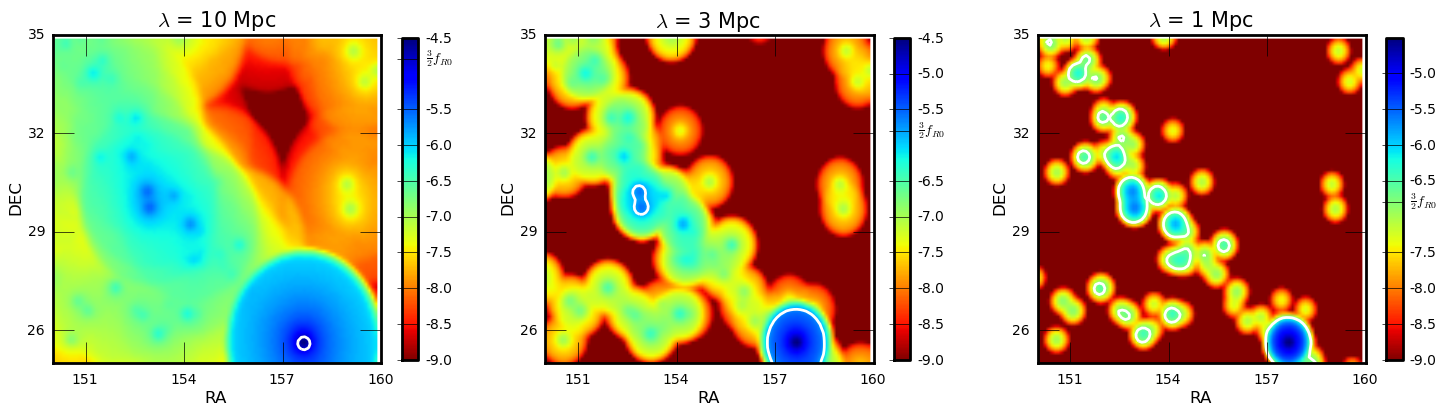}
\caption{The environmental screening map generated in the SDSS region at 210 Mpc ($z=0.05$) for $10 \times 10$ degrees of survey area, approximately $38 \times 38$ Mpc, with 2 arcmin resolution. The external potential $|\phi_{\rm ext}|/c^2$ is shown,  with the screening condition  evaluated using Eq. \ref{eq:criterion22} for models with $f_{R0} = 10^{-5}$, $10^{-6}$ and $10^{-7}$ (left, middle and right panels).  The cut in screening classification is  indicated in the colorbar as $3/2\ f_{R0}$ and also shown in the maps as a white contour line (regions inside the contour are screened). Almost the entire volume is unscreened in the $F5$ model. Even in the $F6$ and $F7$ models most of the volume is unscreened. But since galaxies are preferentially located in or near the high density (screened) regions, a careful evaluation of 
screening for a desired galaxy sample is necessary. 
}
\label{fig:sdss-map}
\end{center}
\end{figure}

We chose a volume that extends to redshift $z=0.05$, which is at a comoving distance of 210 Mpc and has an angular size of 10,000 square degrees. A 10 $\times$ 10 degrees (38 $\times$ 38 Mpc) patch of our map is shown in Fig. \ref{fig:sdss-map}. The map is generated for different values of background Compton length $\lambda_C$ (1 to 10 Mpc); each one corresponds to models with different $f_{R0}$ parameter. 
Fig. \ref{fig:sdss-map} shows the result for the $F5, F6$ and $F7$ models. 
Almost the entire volume  is unscreened under a $F5$ model -- with the exception of massive and rare galaxy clusters, deep enough potential wells do not exist.  While most of the volume also appears unscreened
in the $F6$ and $F7$ models,   galaxies reside preferentially in or near the screened regions so 
one has to be careful -- for a specific galaxy sample, the fraction of screened galaxies depends on their mass distribution since the locations of galaxies (field vs. group) correlate with mass. The screening map and catalogs we make available allow one to 
determine the screening level for a galaxy at any location within the volume covered. 

\begin{figure}[htb]
\begin{center}
  \includegraphics[scale=0.15]{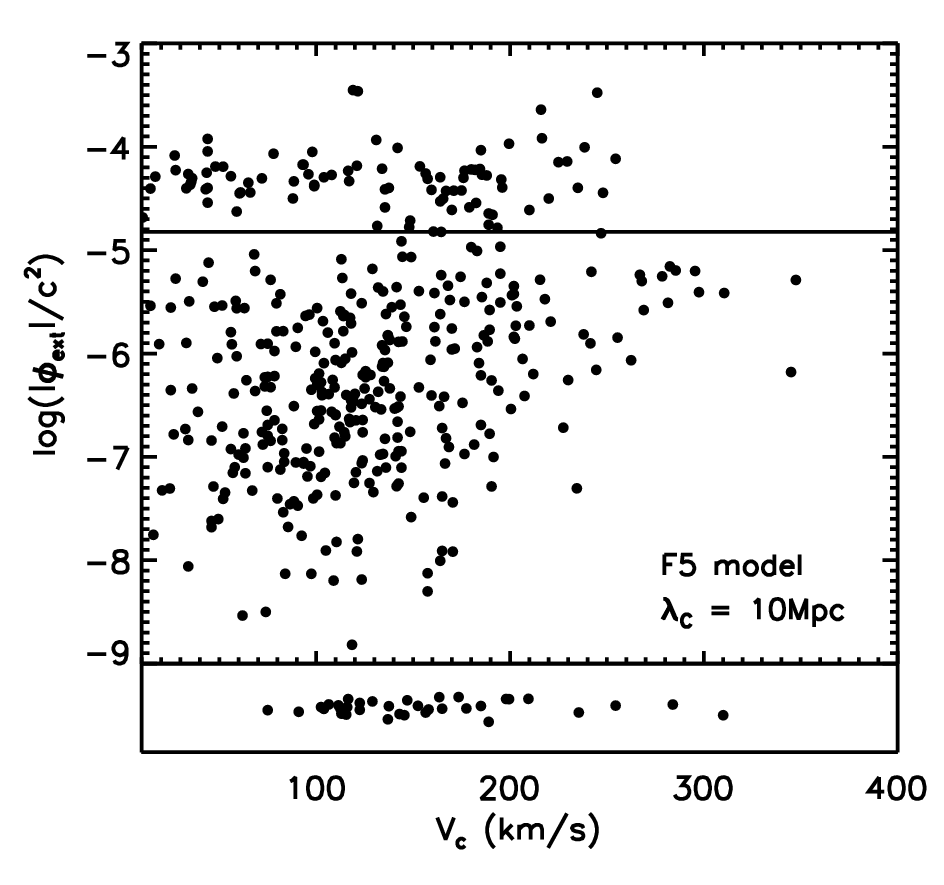}
  \includegraphics[scale=0.15]{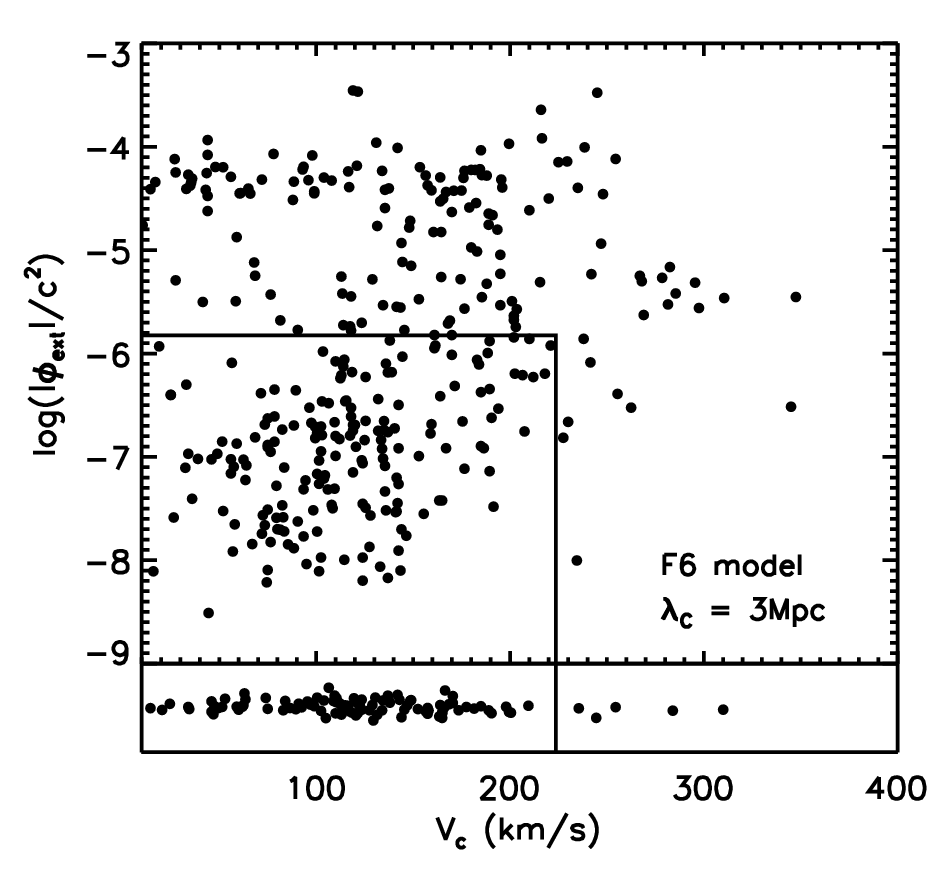}
  \includegraphics[scale=0.15]{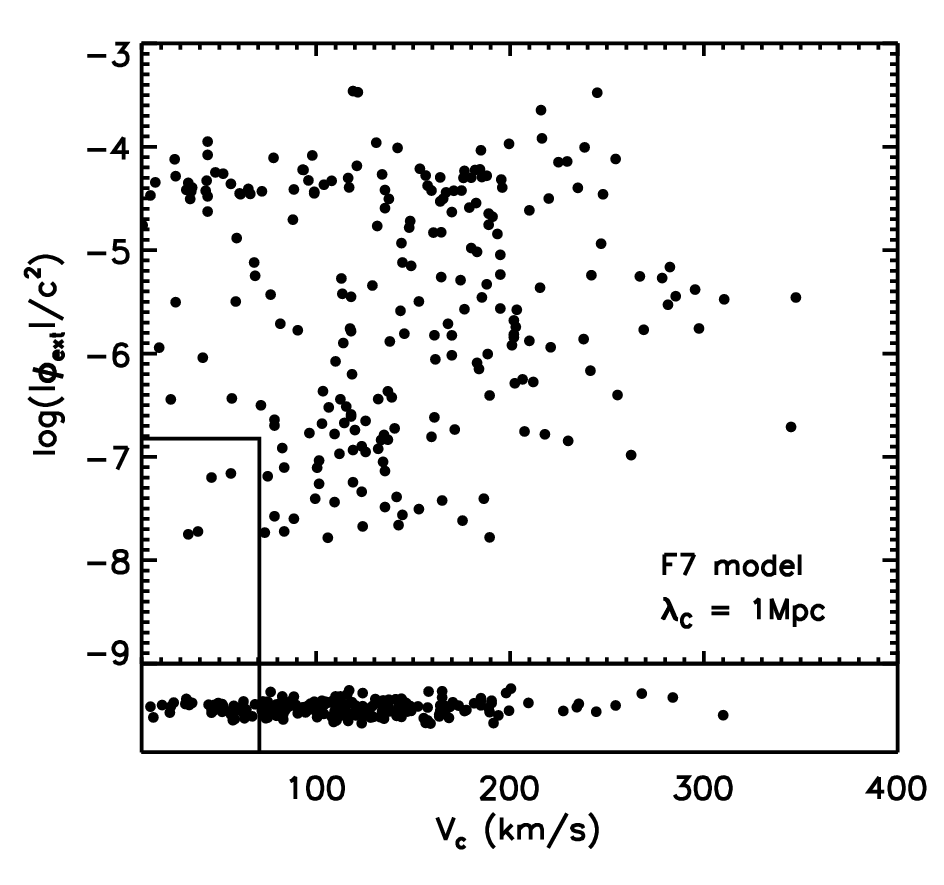}
\caption{We plot $|\phi_{\rm ext}|-v_c$ for SDSS galaxies, for the $f(R)$ models $F5$ (left), $F6$ (middle) or $F7$ (right). 
Zero values (no neighbors within $\lambda_C$) are plotted on the bottom of each panel of the figure.}
\label{fig:plotwarp}
\end{center}
\end{figure}

As an example, we study the scatter in $|\phi_{\rm ext}|-M_{\rm dyn}$ for a set of nearly edge-on, rotationally supported galaxies selected to study warps as a probe for modified gravity following \cite{Jain:2011ji}. 
The details of the test will be presented elsewhere.  These galaxies are selected inside the SDSS footprint. In Fig. \ref{fig:plotcut} we plot $|\phi_{\rm ext}|-v_c$ for these galaxies for models $F5$ (left), $F6$ (middle) and $F7$ (right). The peak  circular velocity $v_c$ is used a proxy for $M_{\rm dyn}$. 
We overplot the limits for unscreened/screened following Eq. \ref{eq:criterion} using the conversion 
\begin{equation}
\left(\frac{v_c {\rm (km/s)}}{100}\right)^2=\frac{f_{R0}}{2\times 10^{-7}} .
\end{equation} 
Note how Fig. \ref{fig:plotwarp} and Fig. \ref{fig:plotphimass} are similar in range since the halo mass limit is similar in data and simulations ($M>5 \times 10^{11}M_\odot$). In the $F5$ model, we suffer from the small number of screened galaxies, but the classification should  still be accurate. The $F6$ model is the most complete while the $F7$ model is likely to have incompleteness issues at larger distances. 

For $f_{R0} < 10^{-6}$ care must be exercised in using our map. Over the full 200 Mpc line of sight distance we do not have completeness for galaxies with halo mass below $10^{12} M_\odot$. Within 10 Mpc we do have information on galaxies with masses well below this limit. At distances beyond that, the mass limit of the combined dataset we have used must be evaluated with care. In particular, the catalogs described above are incomplete over the distance range 10-50 Mpc where the SDSS galaxy catalog should be used directly. In the tests we have carried out, we supplemented the catalogs described here with NED-based investigations of galaxy neighbors.

\section{Conclusions}

In order to test modified gravity theories with screening mechanisms,
we need a criterion to separate screened from unscreened
galaxies that is applicable to galaxy catalogs.
We have tested two simple criteria for chameleon theories with
simulations of $f(R)$ gravity. While the criteria use simplified
approximations to test for screening, we find they are successful at
obtaining screened and unscreened subsamples that are robust to systematics.

For the criterion based on estimation of the external Newtonian potential, given in Eq. \ref{eq:criterion}, we find that
the intrinsic contamination is around 10\%. We model the main expected 
systematics -- mass uncertainty,
incompleteness,  and distance errors -- and
study how the contamination in the method degrades. For pessimistic
assumptions, the contamination increases to about 20\%.
We are able to test the $f(R)$ models $F5$ and $F6$ 
with catalogs from simulations. Our tests of the $F7$ model are limited
by the mass resolution of our simulations but we expect the criteria to scale 
simply to any relevant value of the background fields. 

We  applied our method to a catalog of galaxies in the SDSS region
and constructed a screening map. For  three discrete choices of the 
background field value $f_{R0}$ we have shown examples of screening maps and
 samples of screened and
unscreened galaxies.  
We have used our screening results in two companion studies of gravity tests 
with cepheids and with late-type dwarf
galaxies (Jain et al 2012; Vikram et al 2012). 
We have also made available our compilation of input catalogs and 
code to evaluate screening\footnote{\url{http://www.sas.upenn.edu/~vinu/screening/}}. 

There are several caveats and possible improvements for future work. We have used
approximate screening criteria, but for a given mass distribution the scalar field 
equations can be solved numerically to obtain more accurate screening maps. 
The contamination in the presence of systematics can be tested and controlled by 
discarding galaxies that are expected to be partially screened. Further, for a given
screening level the thin shell condition allows for a quantitative estimate of partial 
screening. Finally, while our results are tested only with $f(R)$ simulations, 
extension to other chameleon theories and symmetron or environmental dilaton 
based screening are feasible -- see e.g. the discussion in Jain et al (2012). 

\section{Acknowledgements}

We are very grateful to Baojiu Li for allowing us to use the halo catalog for the F4, F5 and F6 models, and for his help on obtaining the catalog for the F7 model.  We also thank Xiaohu Yang for sharing the SDSS DR7 group catalogs. We acknowledge helpful discussions with Lam Hui, Mike Jarvis and Fabian Schmidt. AC and BJ are partially supported by NSF grant AST-0908027. GBZ and KK are supported by STFC grant ST/H002774/1. KK is also supported by the ERC and Leverhulme trust.


\begin{thebibliography}{99}

\bibitem[\protect\citeauthoryear{Abell, Corwin and Olowin}{1989}]{abe89}
G.~O.~Abell, Jr.~H.~G.~Corwin and R.~P.~Olowin,
ApJs, 70,  1-138 (1989)

\bibitem[\protect\citeauthoryear{Brax et al}{2010}]{brax2010}
P.~Brax, C.~van de Bruck, A.~C.~Davis and D.~Shaw,
Phys. Rev. D82, 063519 (2010)

\bibitem[\protect\citeauthoryear{De Felice and Tsujikawa}{2010}]{fR_review2}
A.~De Felice, S.~Tsujikawa,
 Living Rev.\ Rel.\  13, 3 (2010) 

\bibitem[\protect\citeauthoryear{Davis et al}{2011}]{davis:2011}
 A.~C.~Davis, E.~A.~Lim, J.~Sakstein and D.~Shaw
 arXiv:1102.5278 [astro-ph.CO] (2011)

\bibitem[\protect\citeauthoryear{Ebeling et al}{1996}]{ebe96}
H.~Ebeling, W.~Voges, H.~Bohringer, A.~C.~Edge, J.~P.~Huchra and U.~G.~Briel, 
MNRAS, 281, 799-829  (1996)

\bibitem[\protect\citeauthoryear{Evrard et al}{2008}]{evr08}
A.~E.~Evrard et al, 
ApJ, 672, 122-137 (2008)


\bibitem[\protect\citeauthoryear{Haas et al}{2011}]{Haas:2011mt}
 M.~R.~Haas, J.~Schaye and A.~Jeeson-Daniel,
MNRAS, 419, 2133-2146 (2012)

\bibitem[\protect\citeauthoryear{Hinterbichler and Khoury}{2010}]{Hinterbichler:2010} 
K.~Hinterbichler and J.~Khoury, 
Phys. Rev. Let., 104, 231301 (2010)


\bibitem[\protect\citeauthoryear{Hu and Sawicki}{2007}]{Hu:2007nk}
 W.~Hu, I.~Sawicki,
 Phys.\ Rev.\  D76, 064004 (2007)
 
\bibitem[\protect\citeauthoryear{Hui, Nicolis and Stubbs}{2009}]{Hui:2009kc}
 L.~Hui, A.~Nicolis, C.~Stubbs,
 Phys.\ Rev.\  D80, 104002 (2009)
 
 
\bibitem[\protect\citeauthoryear{Jain and Khoury}{2010}]{Jain:2010ka}
 B.~Jain, J.~Khoury,
 Annals Phys.\  325, 1479-1516 (2010)
  
\bibitem[\protect\citeauthoryear{Jain}{2011}]{Jain:2011fv}
 B.~Jain, 
Royal Society of London Philosophical Transactions Series A, 369, 5081-5089 (2011)

\bibitem[\protect\citeauthoryear{Jain and VanderPlas}{2011}]{Jain:2011ji} 
B.~Jain and J.~VanderPlas, 
JCAP, 10, 32 (2011)

\bibitem[\protect\citeauthoryear{Jain, Vikram and Sakstein}{2012}]{Jain:2012}
B.~Jain, V.~Vikram and J.~Sakstein,
to be submitted

\bibitem[\protect\citeauthoryear{Karachentsev et al}{2004}]{kar04}
I.~D.~Karachentsev, V.~E.~Karachentseva, W.~K.~Huchtmeier and D.~I.~Makarov
AJ, 127, 2031-2068 (2004)


\bibitem[\protect\citeauthoryear{Khoury and Weltman}{2004}]{Khoury:2003rn}
 J.~Khoury and A.~Weltman,
 Phys.\ Rev.\  D69, 044026 (2004) 

\bibitem[\protect\citeauthoryear{Lavaux and Hudson}{2011}]{lav011}
G.~Lavaux and M.~J.~Hudson, 
MNRAS, 416, 2840-2856 (2011)

\bibitem[\protect\citeauthoryear{Li and Barrow}{2007}]{Li:2007}
 B.~Li and J.~D.~Barrow,
 Phys.\ Rev.\  D75, 084010 (2007)

\bibitem[\protect\citeauthoryear{Li and Zhao}{2009}]{Li:2009}
 B.~Li and H.~Zhao,
 Phys.\ Rev.\  D80, 044027 (2009)

\bibitem[\protect\citeauthoryear{Li and Zhao}{2010}]{Li:2010}
 B.~Li and H.~Zhao,
 Phys.\ Rev.\  D81, 104047 (2010)

\bibitem[\protect\citeauthoryear{Li and Barrow}{2011}]{Li:2011}
 B.~Li and J.~D.~Barrow,
 Phys.\ Rev.\  D83, 024007 (2011) 

\bibitem[\protect\citeauthoryear{Lombriser et al}{2010}]{Lombriser2010}
L.~Lombriser, A.~Slosar, U.~Seljak and W.~Hu,
 arXiv:1003.3009 [astro-ph.CO] (2010)


\bibitem[\protect\citeauthoryear{Reiprich and B{\"o}hringer}{2002}]{rei02}
T.~H.~Reiprich and H.~B{\"o}hringer, 
ApJ, 567, 716-740 (2002)

\bibitem[\protect\citeauthoryear{Schmidt et al}{2009}]{Oyaizu:2008sr}
 F.~Schmidt, M.~V.~Lima, H.~Oyaizu, W.~Hu,
 Phys.\ Rev.\  D79, 083518 (2009)  
 

\bibitem[\protect\citeauthoryear{Schmidt, Vikhlinin and Hu}{2009}]{schmidt:2009}
F.~Schmidt, A.~Vikhlinin and W.~Hu,
Phys.\ Rev.\  D80, 8, 083505 (2009)

 
\bibitem[\protect\citeauthoryear{Schmidt}{2010}]{Schmidt:2010jr}
 F.~Schmidt,
 Phys.\ Rev.\  D81, 103002 (2010) 

\bibitem[\protect\citeauthoryear{Silvestri and Trodden}{2009}]{Silvestri:2009}
A.~Silvestri and M.~J.~Trodden,
Rept.\ Prog.\ Phys., 72, 096901 (2009)

 
\bibitem[\protect\citeauthoryear{Smith}{2009}]{Smith:2009fn}
 T.~L.~Smith,
 arXiv:0907.4829 [astro-ph.CO] (2009)

\bibitem[\protect\citeauthoryear{Song and Koyama}{2009}]{Song:2008vm}
 Y.~-S.~Song, K.~Koyama,
 JCAP 0901, 048 (2009)


\bibitem[\protect\citeauthoryear{Sotiriou and Faraoni}{2010}]{fR_review1}
 T.~P.~Sotiriou, V.~Faraoni,
 Rev.\ Mod.\ Phys.\ 82,  451-497 (2010)

\bibitem[\protect\citeauthoryear{Starobinsky}{2007}]{starobinsky2007}
A.~A.~Starobinsky, 
Soviet Journal of Experimental and Theoretical Physics Letters, 86, 157-163 (2007)


\bibitem[\protect\citeauthoryear{Vainshtein}{1972}]{Vainshtein:1972sx}
 A.~I.~Vainshtein,
 Phys.\ Lett.\ B39, 393-394 (1972)

\bibitem[\protect\citeauthoryear{Vikram et al}{2012}]{Vinu}
V.~Vikram, A.~Cabr\'e, B.~Jain and J.~VanderPlas, in preparation.

\bibitem[\protect\citeauthoryear{Will}{2006}]{will2006}
C.~M.~Will,
Living Reviews in Relativity, 9, 3 (2006)
 
\bibitem[\protect\citeauthoryear{Yang et al}{2005}]{yang:2005}
 X.~Yang, H.~J.~Mo, F.~C.~van den Bosch and Y.~P.~Jing
 MNRAS, 256, 1293 (2005)

\bibitem[\protect\citeauthoryear{Yang et al}{2007}]{yang:2007}
 X.~Yang, H.~J.~Mo, F.~C.~van den Bosch, A.~Pasquali, C.~Li and M.~Barden
 ApJ, 671, 153 (2007)

\bibitem[\protect\citeauthoryear{Zhao et al}{2011a}]{Zhao:2010qy} 
G.~B.~Zhao, B.~Li and K.~Koyama, 
Phys.\ Rev.\  D83, 044007 (2011a) 

\bibitem[\protect\citeauthoryear{Zhao et al}{2011b}]{Zhao:2011cu} 
G.~B.~Zhao, B.~Li and K.~Koyama, 
Phys.\ Rev.\ Lett.\ 107, 071303 (2011b)

 
\end{thebibliography}
\end{document}